\documentclass[aps,prl,groupedaddress,superscriptaddress,reprint,floatfix]{revtex4-1}
\usepackage{graphicx,amsmath,amssymb,dsfont}

\usepackage{bbm}

\newcommand{\vdc}{\ensuremath{V_{\text{dc}}}}
\newcommand{\vsd}{\ensuremath{V_{\text{SD}}}}
\newcommand{\icpt}{\ensuremath{I_{\text{CPT}}}}

\newcommand{\fo}{\ensuremath{f_{0}}}

\newcommand{\ngt}{\ensuremath{n_{g}}}
\newcommand{\nph}{\ensuremath{n_{\text{ph}}}}
\newcommand{\nth}{\ensuremath{n_{\text{th}}}}
\newcommand{\ie}{\textit{i.~e.}}

\newcommand{\rk}{\ensuremath{R_{K}}}
\newcommand{\rn}{\ensuremath{R_{n}}}

\newcommand{\e}[1]{\ensuremath{\times 10^{#1}}}
\newcommand{\units}[1]{\ensuremath{\mathrm{#1}}}
\newcommand{\amount}[2]{\ensuremath{#1\:\units{#2}}}
\newcommand{\cg}{\ensuremath{C_{g}}}
\newcommand{\vg}{\ensuremath{V_{g}}}
\newcommand{\ec}{\ensuremath{E_{c}}}
\newcommand{\csig}{\ensuremath{C_{\Sigma}}}
\newcommand{\ej}{\ensuremath{E_{J}}}
\newcommand{\iv}{$I$-$V$}
\newcommand{\kb}{\ensuremath{k_{B}}}

\newcommand{\zo}{\ensuremath{Z_{0}}}
\newcommand{\zb}{\ensuremath{Z_{b}}}
\newcommand{\wo}{\ensuremath{\omega_{0}}}
\newcommand{\wdr}{\ensuremath{\omega_{d}}}
\newcommand{\vn}{\ensuremath{V_{n}}}
\newcommand{\tn}{\ensuremath{T_{n}}}

\newcommand{\el}{\ensuremath{|0\rangle}}
\newcommand{\eh}{\ensuremath{|1\rangle}}
\newcommand{\zt}{\ensuremath{Z_{t}(\omega)}}

\begin{document}

\title{A Single-Cooper-Pair Josephson Laser}


\author{Fei Chen}
\affiliation{Department of Physics and Astronomy, Dartmouth College, Hanover New Hampshire 03755, USA}

\author{Juliang Li}
\affiliation{Department of Physics and Astronomy, Dartmouth College, Hanover New Hampshire 03755, USA}

\author{A. D. Armour}
\affiliation{School of Physics and Astronomy, University of Nottingham, Nottingham NG7 2RD, United Kingdom}

\author{E. Brahimi}
\affiliation{Department of Physics and Astronomy, Dartmouth College, Hanover New Hampshire 03755, USA}

\author{Joel Stettenheim}
\affiliation{Department of Physics and Astronomy, Dartmouth College, Hanover New Hampshire 03755, USA}

\author{A. J. Sirois}
\affiliation{University of Colorado, 2000 Colorado Ave., Boulder, Colorado 80309, USA}

\author{R. W. Simmonds}
\affiliation{National Institute of Standards and Technology, Boulder, Colorado 80305, USA}

\author{M. P. Blencowe}
\affiliation{Department of Physics and Astronomy, Dartmouth College, Hanover New Hampshire 03755, USA}

\author{A. J. Rimberg}
\email[]{ajrimberg@dartmouth.edu}
\affiliation{Department of Physics and Astronomy, Dartmouth College, Hanover New Hampshire 03755, USA}

\date{\today}

\maketitle

\section{Main Body}

{\bfseries  
The advent of quantum optical techniques based on superconducting circuits\cite{Wallraff:2004} has  opened new regimes in the study of the non-linear interaction of light with matter.   Of particular interest has been the creation of non-classical states of light,\cite{Hofheinz:2009,Sayrin:2011,Kirchmair:2013,Hoi:2012,Brooks:2012}  which are essential for continuous-variable quantum information processing,\cite{Braunstein:2005} and could enable quantum-enhanced measurement sensitivity.\cite{Giovannetti:2004}   Here we demonstrate a device consisting of a superconducting artificial atom, the Cooper pair transistor,\cite{Joyez:1994}  embedded in a superconducting microwave cavity that may offer a path toward simple, continual production of non-classical photons.\cite{Armour:2013}  By applying a dc voltage to the atom, we use the ac Josephson effect to inject photons into the cavity. The backaction of the photons on single-Cooper-pair tunneling events results in a new regime of simultaneous quantum coherent transport of Cooper pairs and microwave photons.  This single-pair Josephson laser offers great potential for the production of amplitude-squeezed photon states and a rich environment for the study of the quantum dynamics of nonlinear systems.\cite{Habib:1998,Chaudhury:2009,Blencowe:2012}
}
\begin{figure}[h]
\includegraphics[width=9cm]{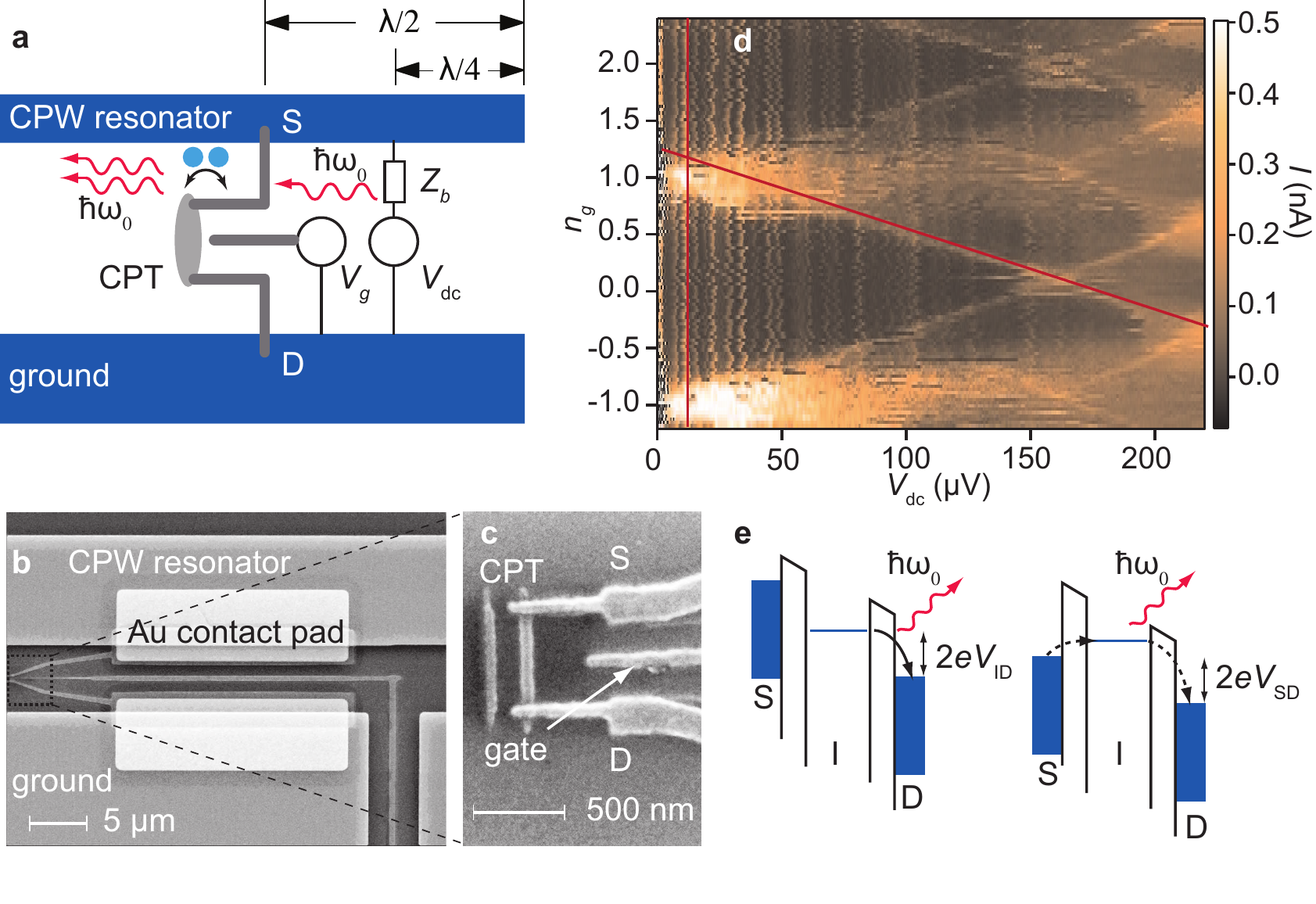} 
\caption{\label{fig1} {\bfseries Single-pair Josephon laser. a,}  Schematic illustration of the circuit. {\bfseries b,} Electron micrograph of the CPT location.  Proximitized \amount{30}{nm} thick Au/Ti contact pads are used to allow electrical contact between the CPT and the cavity without introducing additional dissipation.  The approximately \amount{21.5}{mm}-long cavity is coupled at either end via small capacitors to a waveguide with characteristic impedance $\zo=\amount{50}{\Omega}$.                                                                                    {\bfseries c,} Micrograph of the CPT and its gate line.  The CPT consists of a \amount{7}{nm} thick superconducting Al island with charging energy $\ec=e^2/2\csig= h\times \amount{8.7}{GHz}$, where \csig\ is the total island capacitance and coupled by small Josephson junctions to \amount{70}{nm} thick Al leads.  The Josephson coupling energy $\ej = h\times\amount{17}{GHz}$, which is determined by the junction resistance, is comparable to \ec. {\bfseries d,} Current \icpt\ through the CPT versus \vdc\ and \ngt. {\bfseries e,} Sequential tunneling across the island-drain junction and co-tunneling across the CPT, both with simultaneous net photon emission. For sequential tunneling the energy  released by tunneling across the island-drain junction must match the energy for addition of a photon to the cavity: $2eV_{\text{ID}}=\hbar\wo$.  For cotunneling the source-drain voltage \vsd\ must satisfy a similar condition.  Note that emission of multiple photons is also possible for other bias conditions. }
\end{figure}

The consequences of the ac Josephson effect for electrical transport in Josephson junction systems coupled to microwave fields or electrical resonators have been studied for decades, with the seminal discoveries of such classical phenomena as Shapiro steps\cite{Shapiro:1963} and Fiske modes\cite{Fiske:1964} occurring within only a few years of Josephson's predictions themselves.\cite{Josephson:1962}  This physics was later extended to small junction systems for which electrical transport involves energy exchange between tunneling Cooper pairs and their electromagnetic environment.\cite{Hofheinz:2011}  In the above work, the electromagnetic fields generated by the junctions could generally be considered classical and in the case of small junction systems did not act back on the electrical transport.  One exception relates to recent use of a superconducting charge qubit coupled to a resonator to create a single artificial atom maser.\cite{Rodrigues:2007,Astafiev:2007}  Even here, however, generation of quasiparticles during transport limited the quantum coherence of the overall system.  In contrast, the cavtiy-embedded Cooper pair transistor (cCPT) involves only the interaction of Cooper pairs and photons, in a highly coherent  and heretofore unstudied pumping process that does not depend on dissipative charge relaxation to establish transport.

In our system the CPT is located at the voltage antinode at the center of a wavelength-long coplanar waveguide cavity with a resonant frequency $\wo= 2 \pi \times\amount{5.256}{GHz}$ and quality factor $Q=3.5\e{3}$ giving a photon decay rate $\kappa=2 \pi \times\amount{1.5}{MHz}$.   The cavity, which is fabricated from \amount{100}{nm} thick Nb film, is modified by placement of dc bias lines at the voltage nodes located one quarter wavelength from either end of the cavity, as shown in Fig.~\ref{fig1}a.   These bias lines allow application of a dc bias voltage \vdc\ to the central conductor of the cavity through a biasing impedance \zb\ without affecting the microwave properties of the cavity at its resonant frequency.\cite{Chen:2011}  The CPT itself is fabricated with its source coupled to the central cavity conductor and its drain coupled to the cavity ground plane.  A separate gate voltage \vg\ is applied to the CPT island through a capacitance \cg\ to adjust its electrostatic potential via the gate charge $\ngt = \cg\vg/e$.  Escaping photons can be measured by microwave circuitry connected to the cavity's output port while the dc bias lines are  used to probe electrical transport in the CPT (see Supplementary Information).

For our purposes, the CPT is well described by considering only two charge states, \el\ and \eh, corresponding to zero and one excess Cooper pairs on the island.  These charge states are separated by a gate-dependent electrostatic energy difference $2\varepsilon=4\ec(1-\ngt)$, and coupled to each other via the Josephson energy \ej. By way of the ac Josephson effect, the dc bias gives rise to a characteristic drive frequency $\wdr= e\vsd/\hbar$, which can be viewed approximately as the frequency of Josephson oscillations across each junction.  Here \vsd\ is the source-drain voltage that exists at the CPT\@.  Note that due to the nonlinearity of the system and the presence of the bias impedance $Z_{b}$, \vsd\ in general can differ from the applied voltage \vdc.  Introducing cavity photon annihilation and  creation operators $a$ and $a^{\dagger}$, the Hamiltonian of the cCPT can be expressed as 
\begin{equation}\label{heq}
H = \hbar\wo a^{\dagger}a + \varepsilon \sigma_{z} - \ej \sigma_{x} \cos[\Delta (a + a^{\dagger}) + \wdr t],
\end{equation}
where $\sigma_{x}$ and $\sigma_{z}$ are the Pauli matrices.   The first two terms in (\ref{heq}) describe the cavity photons and the CPT charge.  The third term describes the coupling between the CPT charge states and the cavity photons, and the effects of the voltage drive. In a standard CPT (with no cavity and at zero bias), this term would read $\ej \sigma_{x}\cos\varphi/2$ where $\varphi$, the total superconducting phase difference between the source and drain, can be treated as a classical variable.\cite{Joyez:1994} In our case, however, quantum fluctuations of the cavity photon field must be accounted for via the identification $\hat{\varphi} /2=\Delta (a+a^{\dagger})$, which is proportional to the electric field in the cavity at the location of the CPT.  The dimensionless parameter $\Delta=\sqrt{\zo/\rk}\approx 0.04$, where $\rk=h/e^{2}=\amount{25.8}{k\Omega}$ is the resistance quantum, describes the strength of the quantum phase fluctuations of the cavity field, which can be important for large numbers of photons in the cavity. (See Ref.~\citenum{Blencowe:2012} and the Supplementary Information.  A different possible physical realization of a similar Hamiltonian was also recently studied.\cite{Marthaler:2011})

The current \icpt\ through the CPT measured versus \vdc\ and \ngt\ as in Fig.~\ref{fig1}d shows rich behavior, far more so than in similar measurements of Cooper pair or single electron transistors coupled to lower $Q$ resonators.\cite{Lu:2002,Pashkin:2011}  The current is $2e$ periodic in gate charge for $\vdc\lesssim\amount{150}{\mu V}$, indicating that only Cooper pair transport is significant for sufficiently low bias.  In general, two distinct varieties of transport process attributable to the interaction of the CPT with the cavity are observed.  First, there are sequential tunneling processes involving photon emission, as indicated by the diagonal line in Fig.~\ref{fig1}d.  As shown in the left panel of Fig.~\ref{fig1}e, such processes involve an allowed transition of a Cooper pair across either the source or drain junction, combined with net emission of a photon into the cavity.  Second, there are also higher order virtual processes (cotunneling), indicated by the vertical line in Fig.~\ref{fig1}d.  Here, as shown schematically in the right panel of Fig.~\ref{fig1}e, a Cooper pair is transferred from source to drain through an energetically forbidden state, again with net emission of a photon into the cavity.  It is important to note that in the presence of large numbers of cavity photons, both sequential and co-tunneling can be strongly affected by the cavity field, \ie\ stimulated emission of photons is important, as indicated schematically in Fig.~\ref{fig1}a.

\begin{figure*}[t!]
\includegraphics[width=17cm]{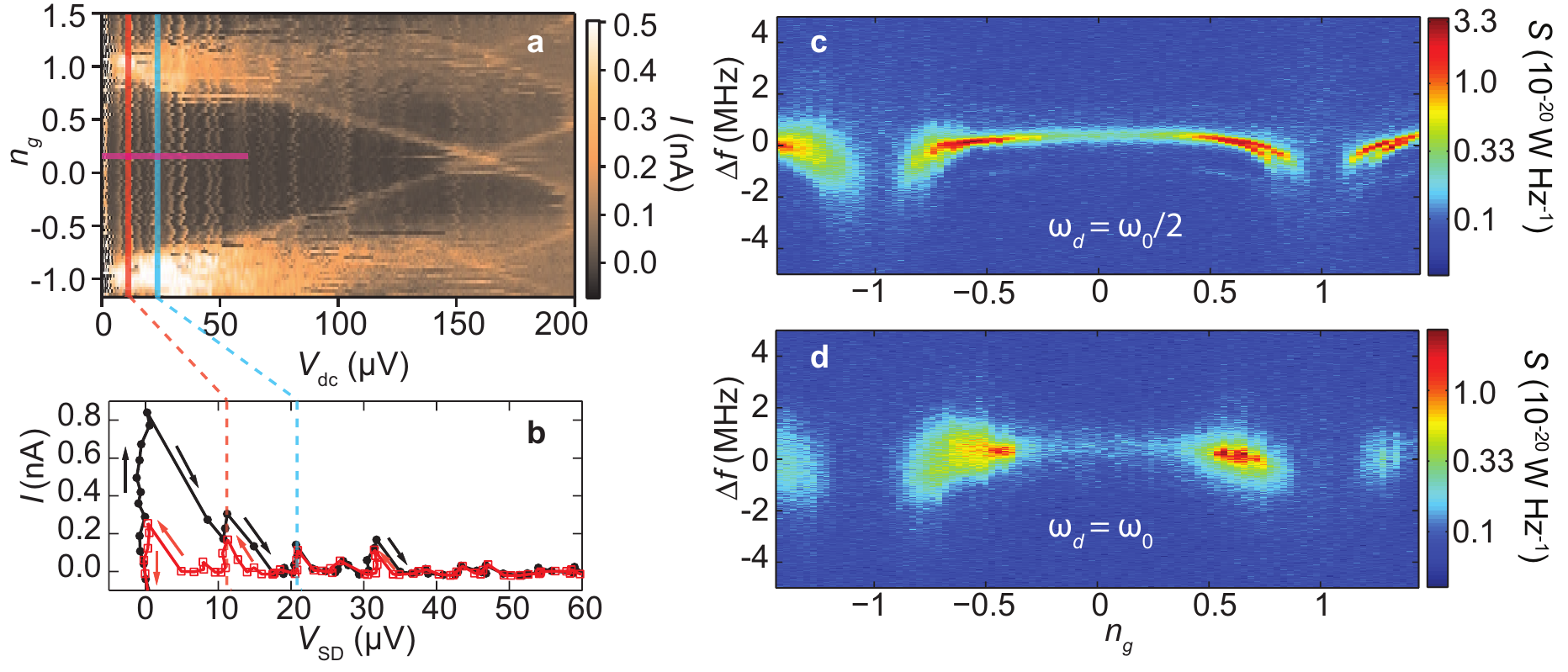} 
\caption{\label{fig2}  {\bfseries Transport and photon emission in the single-pair Josephson laser a,}  Current versus \vdc\ and \ngt\ showing the portions of the parameter space chosen for detailed investigation.  {\bfseries b,} \iv\ characteristic for two sweep directions, as indicated, along the horizontal magenta line in {\bfseries a}.  The vertical dashed red and blue lines indicate the locations of the first and second cotunneling features for $\wdr = \wo/2$ ($\vsd =\amount{11}{\mu V}$) and $\wdr = \wo$ ($\vsd =\amount{22}{\mu V}$) respectively.  {\bfseries c} and {\bfseries d,} Microwave spectral power density $S(\omega)$ of the cCPT over a \amount{5}{MHz} span versus detuning $\Delta f = f-\fo$ from the cavity resonant frequency  $\fo = \wo/2\pi = \amount{5.256}{GHz}$ for the first ({\bfseries c}) and second ({\bfseries d}) cotunneling peaks along the red and blue vertical lines in {\bfseries a}.
}
\end{figure*}

To investigate the coupling to cavity photons more carefully, we restrict ourselves to very low applied voltages ($\vdc<\amount{30}{\mu V}$) and concentrate on the first two cotunneling features in the current as indicated in Fig.~\ref{fig2}a, which occur for $\vsd\approx 11$ and \amount{22}{\mu V}, and correspond to drive frequencies $\wdr \approx \wo/2$ and \wo\ respectively.  Referring to the diagram for cotunneling in Fig.~\ref{fig1}e, these features correspond to the net emission of one or two photons into the cavity during Cooper pair tunneling. Detailed behavior of the CPT current $I$ versus the measured source-drain voltage \vsd\ is shown in Fig.~\ref{fig2}b at a particular gate charge far from the charge degeneracy points at $\ngt=\pm 1$.  The current is hysteretic in \vsd, indicating the presence of bistability in the cCPT dynamics.  Furthermore, there are sharp current steps at fixed voltages \vsd\ corresponding to the cotunneling features.  Such current steps are well known from the phenomena of Shapiro steps\cite{Shapiro:1963} and Fiske modes\cite{Fiske:1964} to be an indication of frequency matching between Josephson oscillations and an electromagnetic field.  In our context, the steps show that the very photons generated by the tunneling Cooper pairs in turn directly influence other tunneling events and subsequent photon emission, \ie, they constitute direct evidence for stimulated photon emission.

To conclusively demonstrate the connection between the current steps in Fig.~\ref{fig2}b and photon emission, we measure the microwave spectral power density $S(\omega)$ emitted by the cCPT for a series of different values of \ngt\ for bias voltages surrounding the cotunneling features near $\vsd \approx\amount{11}{\mu V}$ and \amount{22}{\mu V}.  In order to minimize noise in the dc bias voltage \vdc\ and accompanying jitter in the drive frequency \wdr, for the emission data shown here we used low-noise bias circuitry that significantly improved emission stability (see Supplementary Information).  For both values of \vsd\ there is strong emission close to the cavity resonance frequency \wo\ as can be seen in Fig.~\ref{fig2}c,d. For the first cotunneling feature, at which the drive frequency $\wdr = \wo/2$, emission at twice the drive frequency is a direct consequence of the strong nonlinearity of the system. The emission pattern is $2e$-periodic, as is the electrical transport, and shows interesting structure versus \ngt.  In particular the emission for $\wdr = \wo/2$ appears to develop internal structure for $\ngt\gtrsim 0.6$ while for $\wdr = \wo$ there is a clear ``hot spot'' in the emission for $\ngt \approx 0.7$.  In both cases, the emission dies out as the gate charge approaches the charge degeneracy points at $\ngt=\pm 1$.  

A detailed view of the emission at $\wdr=\wo/2$ and $\wdr=\wo$ reveals  additional interesting features, as shown in Figs.~\ref{fig3} and \ref{fig4} which show the emission spectra $S(\omega)$ and cavity photon occupation $\nph$ versus applied voltage \vdc\ at representative values of gate charge \ngt\ for both the $\wdr =\wo/2$ and $\wdr=\wo$ cotunneling features.  For the $\wdr = \wo/2$ resonance in Fig.~\ref{fig3} we see that as might be expected for a single-atom emitter there is no clear sign of a lasing threshold, with the cavity occupation \nph\ climbing smoothly from zero as \vdc\ is increased.  For low $\vdc\lesssim \amount{13}{\mu V}$ the emission linewidth of roughly \amount{1}{MHz} shows modest narrowing over the intrinsic cavity linewidth $\kappa = \amount{1.5}{M Hz}$.  At $\vdc\approx\amount{13}{\mu V}$, there is a sharp change in the emission pattern: the linewidth suddenly drops by roughly an order of magnitude, to as low as \amount{70}{kHz}, slightly larger than the residual jitter of about \amount{25}{kHz} in the drive frequency $\wdr/2\pi$.  Over the same range in \vdc\ the cavity photon occupancy \nph\ reaches a maximum value on the order of 100 before dropping sharply at $\vdc\approx\amount{13}{\mu V}$, stabilizing briefly, and then rapidly declining.

Strikingly, as charge degeneracy is approached, the sharpened spectrum splits into two narrowly separated peaks at around $\ngt=0.62$.  The separation of these peaks increases as \ngt\ approaches charge degeneracy, accompanied by a notable shift in the emission frequency toward negative detuning.  There is an even more notable shift toward negative detuning as \nph\ increases for fixed \ngt.  This latter tendency results in the characteristic  ``v'' shape of the emission spectra versus \vdc\ as \nph\ first rises and then falls with increasing bias.

The $\wdr = \wo$ emission spectra shown in Fig.~\ref{fig4} share some features with those for $\wdr =\wo/2$; there are, however, significant differences as well.  There is again no clear sign of a threshold as \vdc\ is increased, and there is again a clear though less pronounced pulling toward negative detuning as charge degeneracy is approached or \nph\ is increased.  There is no sudden sharpening of the spectrum in this case; instead, the emission simply cuts off abruptly for $\vdc\gtrsim\amount{23}{\mu V}$, just after the cavity reaches its maximum occupancy of roughly $\nph\approx 200$.  The minimum linewidth of the emission spectra is roughly \amount{500}{kHz}, significantly below the bare cavity linewidth, but not by as much as for the $\wdr=\wo/2$ resonance.   All in all, the combination of linewidth reduction, high cavity photon number, and Shapiro-like \iv\ characteristics provide clear evidence of lasing driven by single-Cooper-pair tunneling.

For many years, the standard theoretical approach for describing such tunneling has been so-called $P(E)$ theory,\cite{Devoret:1990,Girvin:1990} which considers the process of emission or absorption of photons by tunneling Cooper pairs to or from an environment described by a frequency-dependent impedance \zt.  As recent experiments\cite{Hofheinz:2011} on an effective single junction system have shown, if the environmental impedance \zt\ is sharply peaked at  a frequency $\wo$ (as for a resonance) there is substantial probability $P(E)$ of emission of a photon into the environment when the junction is biased at a voltage given by $V=\hbar\wo/2e$.  The result is a peak in the current due to incoherent tunneling of Cooper pairs, combined with emission of microwave photons, similar in some respects to the results presented here.\cite{Hofheinz:2011,Gramich:2013}  

It is important to note, however, that the above picture fails to accurately describe the Josephson laser.  The key point is the assumption essential to $P(E)$ theory that the environment is in thermal equilibrium.  At low temperature $T = \amount{30}{mK}$ and high frequency $\wo = 2 \pi\times \amount{5.26}{GHz}$ the thermal occupation of the environmental modes is very small ($\nth=2\e{-4}$), and the probability $P(E)$ of absorption of a photon by the CPT is essentially zero.  In our experiment, however, the occupation the cavity mode is in fact large ($\nph \approx 100$) and absorption of photons is commonplace, as indicated by the Shapiro-like steps in the \iv\ characteristics of the CPT\@.  Experimentally, this is a consequence of the very large $Q$ of the cavity in our experiment, which ensures that the photons generated by Cooper pair tunneling remain to interact with subsequent tunneling events for long after they are generated. Quantitatively, the product of the Cooper pair tunneling rate $I/2e$ and the photon dwell time of the cavity $2 \pi/\kappa$  is in our case  $\pi I/\kappa e = 480$; for related experiments\cite{Hofheinz:2011} in a regime for which $P(E)$ theory is applicable we estimate $\pi I/\kappa e\approx 2$, over two orders of magnitude smaller.  

Recent experiments on lasing in a related single-electron transistor/cavity system have a photon dwell time comparable to that reported here.\cite{Astafiev:2007}  However, in that case charge transport was based on a combination of Cooper pair and quasiparticle tunneling.  The quasiparticle transitions served to establish population inversion of a three level system similar to that of single atom lasers.\cite{McKeever:2003} The typical photon emission rate for this artificial superconducting atom of roughly \amount{200}{MHz} was small compared to the quasiparticle tunneling rate, typically 3--\amount{5}{GHz}.  As a result, any interactions between tunneling Cooper pairs and the cavity photons were incoherent.  In contrast, for the single-Cooper-pair Josephson laser transport occurs via cotunneling of Cooper pairs, a very different transport process not clearly analogous to single-atom lasing.  Furthermore, the decoherence-inducing influences of the electromagnetic environment and quasiparticle tunneling are vastly reduced relative to previous results.  The result is a new regime of hybrid electronic/photonic transport that cannot be adequately treated within the framework of conventional $P(E)$ theory.
 
Some aspects of this behavior can be explained within a semiclassical model of the cCPT system (see Supplementary Information for details).   The emission of cavity photons at \wo\ for both $\wdr = \wo/2$ and \wo\ resonances can be explained dynamically by considering the time dependence of the drive term $\sigma_{x} \cos[\Delta (a + a^{\dagger}) + \wdr t]$ in the cCPT Hamiltonian.   It can be shown that $\sigma_{x}$ oscillates at odd harmonics of the drive frequency \wdr; the nonlinearity of the system in the form of the product of $\sigma_{x}$ with a sinusoidal function of \wdr\ leads to an overall oscillation at the cavity resonant frequency \wo.  The semiclassical analysis also correctly predicts the approximate number of photons \nph\ in the cavity for both cases, as well as the vanishing of the emission at the charge degeneracy points.  Nonetheless, many aspects of the emission remain unexplained, making the cCPT emission a rich topic for continued theoretical investigation.

\begin{figure}[t!]
\includegraphics[width=9cm]{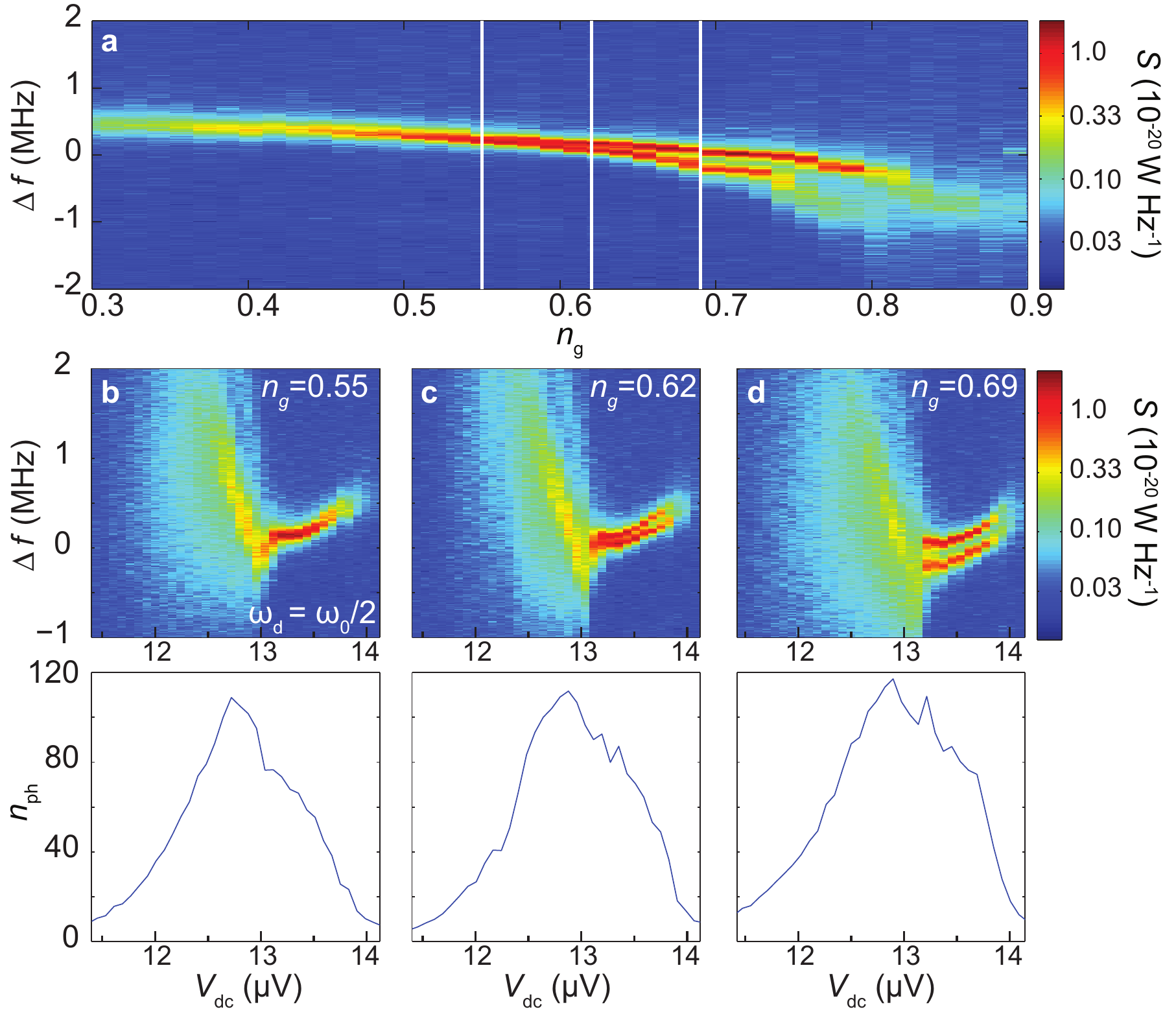} 
\caption{\label{fig3} {\bfseries Detailed emission spectra for $\wdr \approx \wo/2$ . a,}  Emission versus gate charge \ngt\ and detuning $\Delta f$ at close to maximum emission.  {\bfseries b--d,} Top panels: emission versus applied bias voltage \vdc\ and detuning $\Delta f$ for $\ngt=0.55$, 0.62 and 0.69, as indicated by the white vertical lines in {\bfseries a}.  \vdc\ was swept from low to high bias in \amount{80}{nV} increments. Bottom panels: cavity photon occupation $\nph = P/\kappa\hbar\wo$ versus \vdc\ for each gate voltage where $P$ is the integrated emission power. All emission spectra are plotted versus detuning  $\Delta f$.  }
\end{figure} 

Finally, we turn to the question of the nature of the photon field in the cavity, and whether it is best described as a quantum state.  A classical electrical current is well known (ignoring fluctuations) to produce a coherent state of the electromagnetic field.  Given the highly quantum nature of the electronic/photonic transport in the cCPT it seems likely that the cavity photon field has non-classical correlations.    By analogy with single-emitter lasers, which lack the noise associated with multiple emitters and spontaneous emission processes,\cite{Haroche:2006,McKeever:2003} we expect that the cavity photons will be number squeezed; \ie, the cavity photon statistics will be sub-Poissonian, with a Fano factor $F = (\langle \nph^{2}\rangle -\langle\nph\rangle^{2})/\langle\nph\rangle$ less than unity.  

This expectation is supported by several theoretical calculations.\cite{Armour:2013}  Far from the charge degeneracy points we may treat the CPT as an effective single junction system.  Making this approximation and using a rotating wave approximation, we find that the predicted Fano factor for system parameters comparable to the experiment is in the ranges $F\approx 0.6$--0.8, which would place the system firmly in the quantum regime for which  $F<1$.  In fact, suppression of photon number fluctuations leading to a reduced Fano factor seems to be a generic property of the system, whether biased on the cotunneling features considered here or on the sequential tunneling features\cite{Marthaler:2011} also visible in Fig.~\ref{fig1}d.  (See Supplementary Information for details.)  Future experimental studies will focus on use of recently developed state reconstruction techniques using linear detectors\cite{Eichler:2011,Eichler:2012} to accurately characterize the quantum state of photons generated by the Josephson laser.

\begin{figure*}[t!]
\includegraphics[width=12cm]{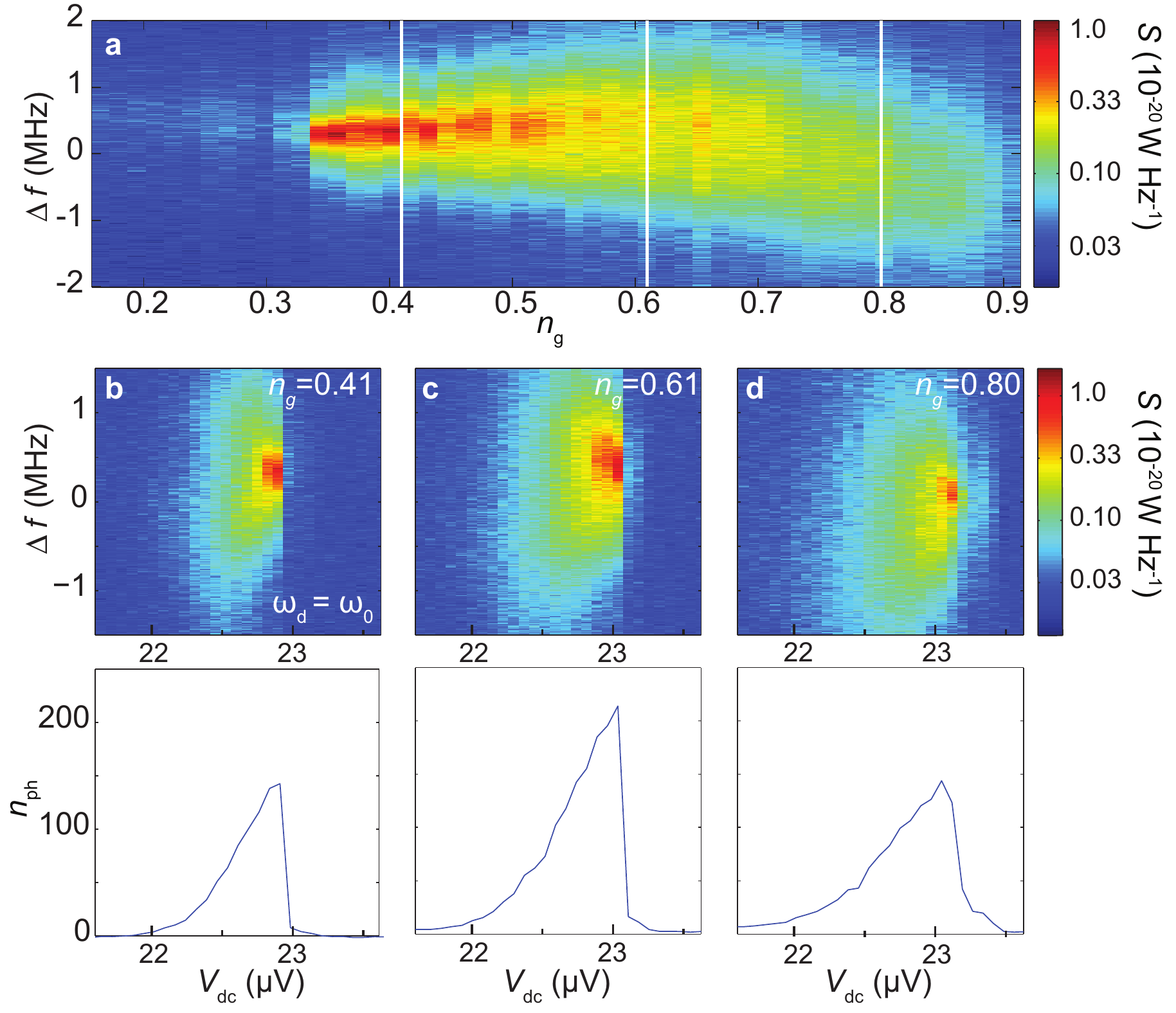} 
\caption{\label{fig4}  {
\bfseries Detailed emission spectra for $\wdr \approx \wo$ . a,}  Emission versus gate charge \ngt\ at close to maximum emission.  {\bfseries b--d,} Top panels: emission versus applied bias voltage \vdc\ for $\ngt=0.41$, 0.61 and 0.80, as indicated by the vertical white lines in {\bfseries a}.  \vdc\ was swept from low to high bias in \amount{75}{nV} increments. Bottom panels: cavity photon occupation $\nph = P/\kappa\hbar\wo$ versus \vdc\ for each gate voltage where $P$ is the integrated emission power. All emission spectra are plotted versus detuning  $\Delta f$.   All emission spectra are plotted versus detuning  $\Delta f$.  }
\end{figure*}

In conclusion, we have demonstrated lasing by means of a new quantum coherent transport process involving the interaction of Cooper pairs and photons.  The single-Cooper-pair Josephson laser may ultimately serve as a convenient, easy-to-use source of amplitude squeezed light, and could form the basis of a new class of electrical or photonic amplifiers.  It could also serve as an important platform for the study of the quantum dynamics of strongly nonlinear systems.  

\section*{Acknowledgements}
This work was supported by the NSF under grants DMR-1104790 and DMR-1104821, by AFOSR/DARPA under agreement FA8750-12-2-0339, and by EPSRC (U.K.) under Grant No. EP/I017828/1.

\section*{Author Contributions}

F.C. and  J.L. fabricated the samples.  F.C. performed the measurements with assistance from J.L. and J.S\@.  A.D.A., M.P.B. and E.B. developed the theoretical model. A.J.S. and R.W.S. fabricated the coplanar waveguide resonators. A.J.R., M.P.B. and A.D.A. conceived and designed the experiment, and co-wrote the manuscript with input from all authors.  

\section*{Competing Financial Interests}

The authors acknowledge no competing financial interests.


\newpage

\renewcommand{\thefigure}{S\arabic{figure}}
\renewcommand{\theequation}{S\arabic{equation}}

\setcounter{section}{0}
\setcounter{figure}{0}

\section*{Supplementary Information}

In the following we provide additional details regarding several aspects of the experiment and associated theoretical investigation.  We begin with a discussion of the techniques used to produce the resonators and the embedded CPTs, and of how important CPT parameters such as charging and Josephson energies are determined.  We then discuss the two biasing schemes used to produce the experimental results in the manuscript.  One biasing scheme was used for low-noise current measurements, while the other was used to produce highly stable emission for spectral measurements.  Next, we turn to the techniques used for measuring cavity emission, and for characterization and modeling of the system amplifier noise.   Finally, we conclude with a more in-depth discussion of our current theoretical understanding of the system.  

\subsection*{Sample Fabrication and Characterization}

\begin{figure}[h]
\includegraphics[width=9cm]{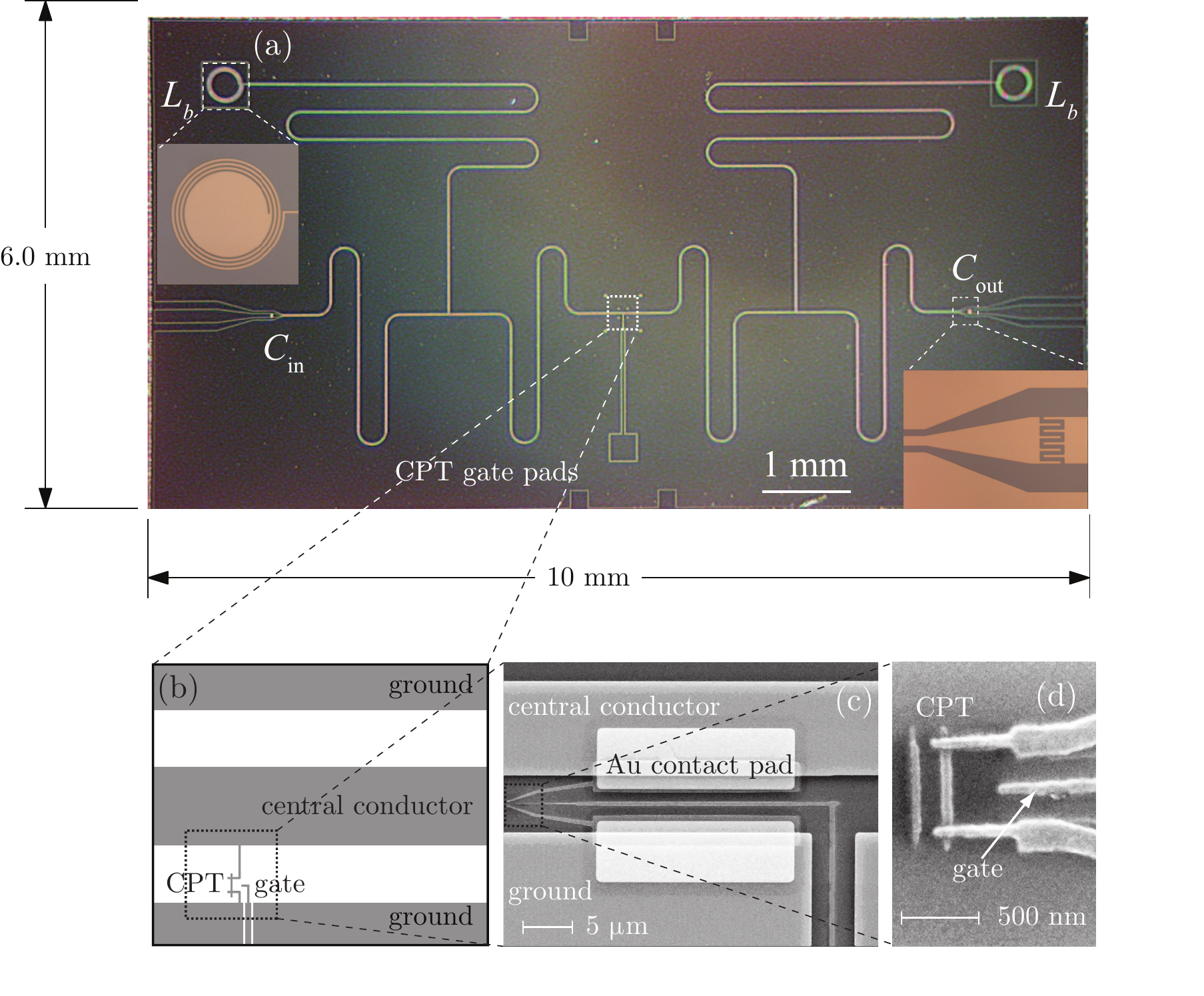} 
\caption{\label{figs1} {\bfseries Sample design.  a,} Optical micrograph of a dc-biased cavity showing the location of the CPT and typical designs for an output capacitor and dc bias line inductor. {\bfseries b,} Schematic illustration of the location of the CPT relative to the cavity. {\bfseries c,} Electron micrograph of the coupling between the CPT and the contact pads. {\bfseries d,} Electron micrograph of a typical CPT\@.  }
\end{figure}

The cavities used for fabricating our cCPTs were based on coplanar waveguide  with inductively terminated dc bias lines attached to the main cavity line a distance $\lambda/4$ from the ends of the cavity, as described in detail elsewhere.\cite{Chen:2011}  The Nb film of the cavity was \amount{100}{nm} thick and was etched with sloping side walls to allow for good step coverage during deposition of the CPT\@.  Oxide was removed under the Au/Ti contact pads by ion milling before deposition of the metal.  A micrograph of a typical cavity is shown in Fig.~\ref{figs1}a.

The central cavity conductor was \amount{10}{\mu m} wide with \amount{5.5}{\mu m} gaps between the central conductor and ground.  Capacitive coupling to the cavity was strongly asymmetric, with a small (\amount{1.7}{fF}) input capacitor and a large (\amount{18.5}{fF}) output capacitor (right inset, Fig.~\ref{figs1}a).   The inductors on the dc bias lines (left inset, Fig.~\ref{figs1}a) were large enough (\amount{5.8}{nH}) that few photons escape from the cavity through the bias circuitry.  Overall, we estimate that roughly 95\% of the cavity photons are collected by the microwave circuitry for amplification and subsequent analysis.   The measured cavity $Q$ of 3500 is in good agreement with the simulated $Q$ (including the dc bias lines) of 3600.  

The CPT itself was fabricated in a separate step using standard electron beam lithography and shadow evaporation techniques, \cite{Dolan:1977}  and aligned with the cavity so as to lie in the gap between the central conductor and ground planes as shown schematically in Fig.~\ref{figs1}b.  The leads of the CPT were thick enough (\amount{70}{nm}) to drive the underlying Au/Ti contact pads in Fig.~\ref{figs1}c superconducting by means of the proximity effect.  The CPT island, on the other hand, was very thin (\amount{7}{nm}), and deposited using a cooled evaporation stage to ensure electrical continuity.  This technique increases the superconducting gap of the CPT island, preventing quasiparticle trapping and ensuring $2e$-periodicity of the CPT \iv\ characteristics.\cite{Xue:2009}

Three separate cCPT samples were fabricated in the course of this experiment.  Of these, microwave measurements of cavity output were performed on two.  All three samples showed similar behavior in their \iv\ characteristics, with all  major features described in the main text clearly visible.  Both samples for which microwave measurements were performed showed microwave photon emission at the first two cotunneling features.  The detailed dc and microwave measurements presented  in the main text were all obtained from a single sample.  
\begin{figure*}[ht!]
\begin{center}
\includegraphics[width=12cm]{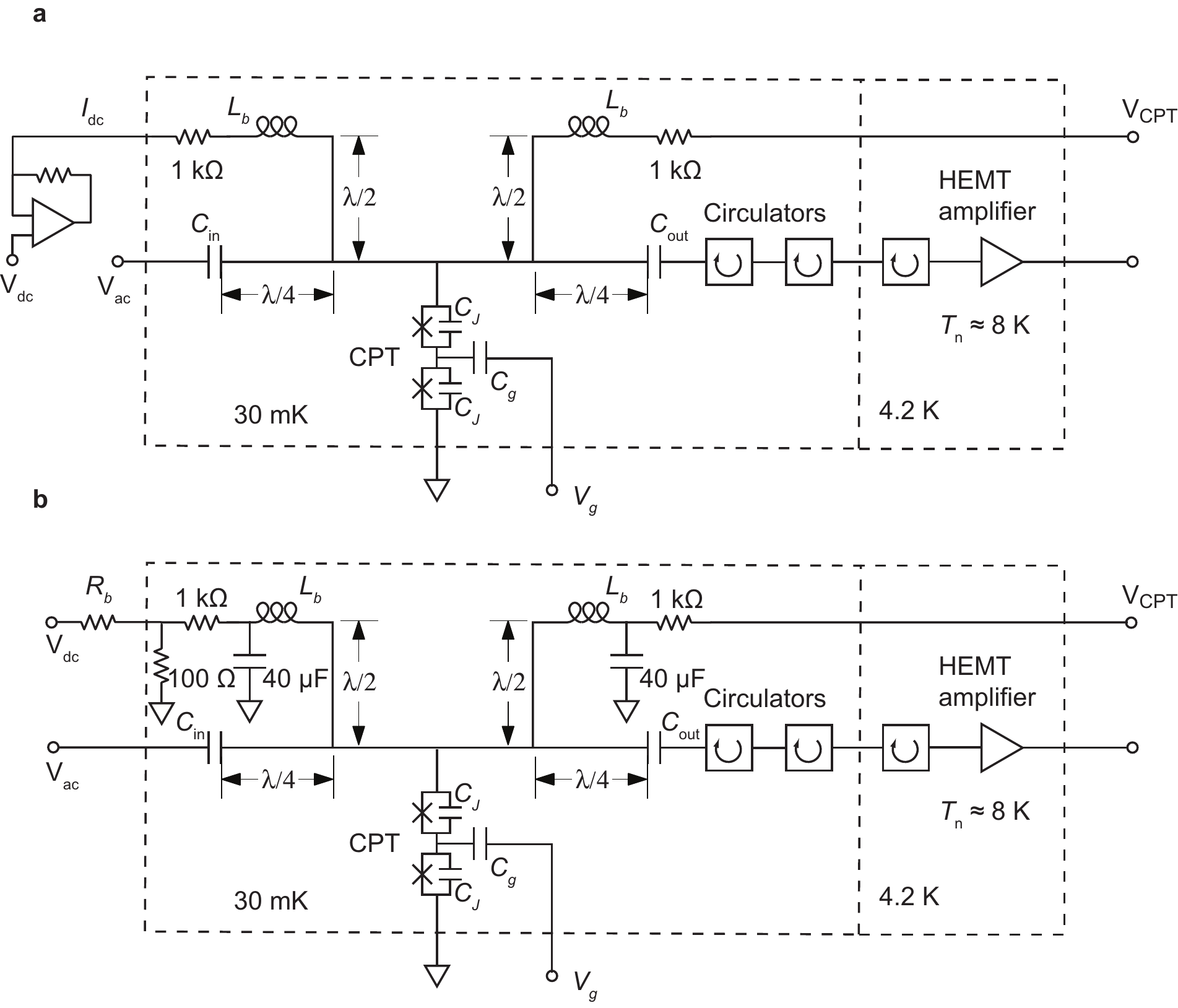} 
\caption{\label{figs2}  {\bfseries  DC and microwave circuitry. a,} Circuit for low-noise current measurements using a floated current sensitive amplifier.  Detection of microwave signals is possible with this setup but emission is not stable.  {\bfseries b,} Circuit for stable microwave emission measurements.  The use of a cold voltage divider and a large capacitor provide excellent bias stability but limit resolution in measurements of current. }
\end{center}
\end{figure*}

The junction and gate capacitances for the sample were determined from the dc transport data.  The $2e$ periodicity of the supercurrent determines the gate capacitance $\cg=\amount{4.6}{aF}$.  The junction capacitances are determined by fitting to the sequential tunneling plus photon emission features described in the main text, giving a source capacitance $C_{S}= \amount{1.08}{fF}$ and a drain capacitance $C_{D}=\amount{1.14}{fF}$.    The resulting total capacitance $\csig = C_{S}+ C_{D} + \cg = \amount{2.22}{fF}$ gives  a charging energy $\ec = e^{2}/2\csig = \amount{36}{\mu eV}=\amount{0.42}{K}$.  The Josephson energy \ej\ was determined from the total normal-state CPT tunneling resistance $\rn=\amount{21.5}{k\Omega}$ through the Ambegaokar-Baratoff relation $\ej=\frac{\Delta}{4}\frac{\rk}{\rn}$, giving $\ej\approx \amount{64}{\mu eV}=\amount{0.74}{K}$. Here $\Delta=\amount{212}{\mu eV}$ is the effective superconducting gap when the different gap sizes of the leads and island are taken into account, and $\rk\approx\amount{25.8}{k\Omega}$ is the resistance quantum.\cite{Xue:2009}

All electrical measurements were performed in a dilution refrigerator at its base temperature of \amount{30}{mK}.  The dc bias lines were filtered with room temperature $RC$ and $\pi$-type filters, as well as cryogenic copper powder microwave filters.  The sample was mounted in a sealed gold-plated Cu box designed to minimize propagation of unwanted microwave modes. The box was in turn located inside a cryogenic Amumetal magnetic shield to minimize the residual magnetic field at the sample location.  

\subsection*{dc Biasing Schemes}

We used two separate dc biasing schemes during the experiment, as shown in Fig.~\ref{figs2}.  One scheme is optimized for sensitive measurements of current and was used for the \iv\ measurements shown in the main text.  The other is optimized for low voltage noise on the bias line and is used for emission  measurements.

In Fig.~\ref{figs2}(a) we show our standard circuitry for low-noise current measurements.  Cold \amount{1}{k\Omega} resistors on the dc bias lines are used to help protect the cCPT from electrostatic discharge.  On the ac input side of the cavity, multiple cold attenuators (total \amount{50}{dB}, not shown) are used to prevent noise from entering the cavity.  On the output side, three circulators are interposed between the cavity and the cryogenic HEMT amplifier to prevent the noise wave at the amplifier input from entering the cavity.  After the HEMT amplifier, additional room temperature amplification is performed on the signal before further measurement.  

Since the cCPT is grounded internally in the dilution refrigerator, the current sensitive amplifier is floated and the dc bias voltage is applied indirectly to the cCPT through the amplifier.  This provides current resolution on the order of a few fA, allowing the highly detailed \iv\ measurements shown in the main text.  

Emission stability is limited when using the measurement circuitry in Fig.~\ref{figs2}a for two reasons.  First, noise associated with the room temperature bias circuitry can cause significant fluctuations in the drive voltage, leading to fluctuations in the drive frequency \wdr. These can be only partially eliminated by use of low-pass filtering at room temperature.  

Further improvement can be made as in Fig.~\ref{figs2}b by adding a cryogenic \amount{100}{\Omega} resistor in parallel with the \amount{1}{k\Omega} protective resistor and sample.  In combination with a room temperature biasing resistor $R_{b}$ the \amount{100}{\Omega} resistor forms a cold voltage divider  that drastically reduces the amplitude of low-frequency noise reaching the sample.  

More fundamentally, however, we must also consider the thermal noise associated with the cold \amount{1}{k\Omega} protective resistor; this noise is not negligible despite the resistor's low temperature.   For instance, assume that thermal noise in a bandwidth $B\approx\amount{1}{MHz}$ from the \amount{1}{k\Omega} resistor can reach the sample.  The associated rms noise voltage $\vn=\sqrt{4 \kb T B} = \amount{40}{nV}$, when applied to the cCPT, results in a drive frequency jitter $\delta \wdr= 2 \pi \times e\vn/h \approx \amount{10}{MHz}$ that is far larger than the cavity bandwidth $\kappa=2\pi\times\amount{1.5}{MHz}$.  

This noise can be significantly reduced, however, by placing a large cryogenic capacitor $C\approx\amount{40}{\mu F}$ between the \amount{1}{k\Omega} resistor and the cCPT\@.  The resulting noise reaching the cCPT can be estimated as $\vn=\sqrt{\kb T/C} \approx \amount{0.1}{nV}$.  The drive frequency jitter is then drastically reduced to $\delta\wdr \approx 2 \pi\times \amount{25}{kHz}$, on the order of the sharpest spectral features we measure.  This vastly improved bias stability has made possible the highly detailed emission and state tomography measurements presented in the main text.

\subsection*{Microwave  Measurements}

After exiting the cavity, microwave photons pass through three circulators and then enter the amplifier chain,  which consists of a cryogenic HEMT amplifier with noise temperature $\tn=\amount{8}{K}$ and gain $G_{1}=\amount{38}{dB}$ followed by a bandpass filter and then a room temperature FET amplifier with gain $G_{2}=\amount{45}{dB}$.   

\begin{figure}[ht!]
\includegraphics[width=7cm]{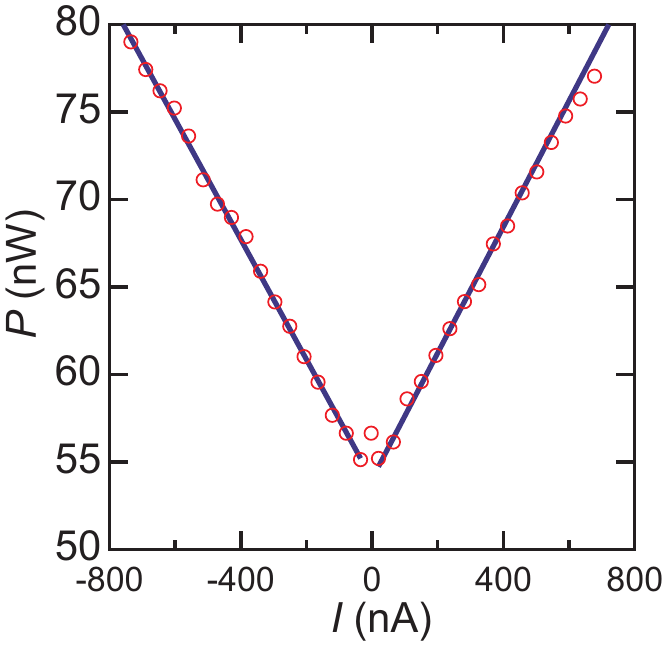} 
\caption{\label{figs4} {\bfseries Amplifier chain noise characterization.  } Noise power emitted by the cCPT in its bandwidth $\kappa$ (red circles) versus driving current $I$.  The slope of the blue fitted lines gives the total system gain, while their intercept gives the system noise temperature.  }
\end{figure}

We can use the shot noise of the CPT itself, when driven on its quasiparticle branch, to fully calibrate the amplifier chain.  This is essentially the same technique long used to calibrate system gain and noise for the related RF-SET\@.\cite{Aassime:2001a}  A straightforward calculation shows that the shot noise power $P_{\text{SN}}$ delivered by the cCPT to the HEMT amplifier is given by 
\begin{equation}\label{sneq}
P_{\text{SN}} =\frac{e I}{4 \mathcal{C}}\frac{Q_{e}}{Q_{c}}
\end{equation}
where $I$ is the driving current and $\mathcal{C}$ is the total cavity capacitance.  The cavity $Q$ is written here as $Q_{c}$, while $Q_{e}$ is the cavity $Q$ when additionally loaded by the CPT high bias tunneling resistance $\rn$, given by $Q_{e}^{-1} = Q_{c}^{-1} + Q_{\text{CPT}}^{-1}$ where $Q_{\text{CPT}}=\wo\rn\mathcal{C}$.  As can be seen in Fig.~\ref{figs4}a, the measured shot noise power clearly scales linearly with the applied driving current.  From (\ref{sneq}), the slope of the measured power $P_{\text{SN}}$ versus $I$ in Fig.~\ref{figs4}a is given by $e G Q_{e}/4 \mathcal{C} Q_{c}$ where $G$ is the gain of the amplifier chain.   The $y$ intercept $P_{0}$ of the linear fits for positive and negative $I$ give the system noise temperature via the relation $\tn = P_{0}/\kb G\,\text{BW}$ where $\text{BW} = \amount{20}{MHz}$ is the bandwidth of the measurement.  The data in Fig.~\ref{figs4}a give a system gain of \amount{68}{dB} at the output of the FET amplifier and a system noise temperature $\tn=\amount{31}{K}$.  

\subsection{Derivation of the CCPT Hamiltonian}
\label{hamiltoniansec}

\begin{figure}
\begin{center}
\includegraphics[width=8.5cm]{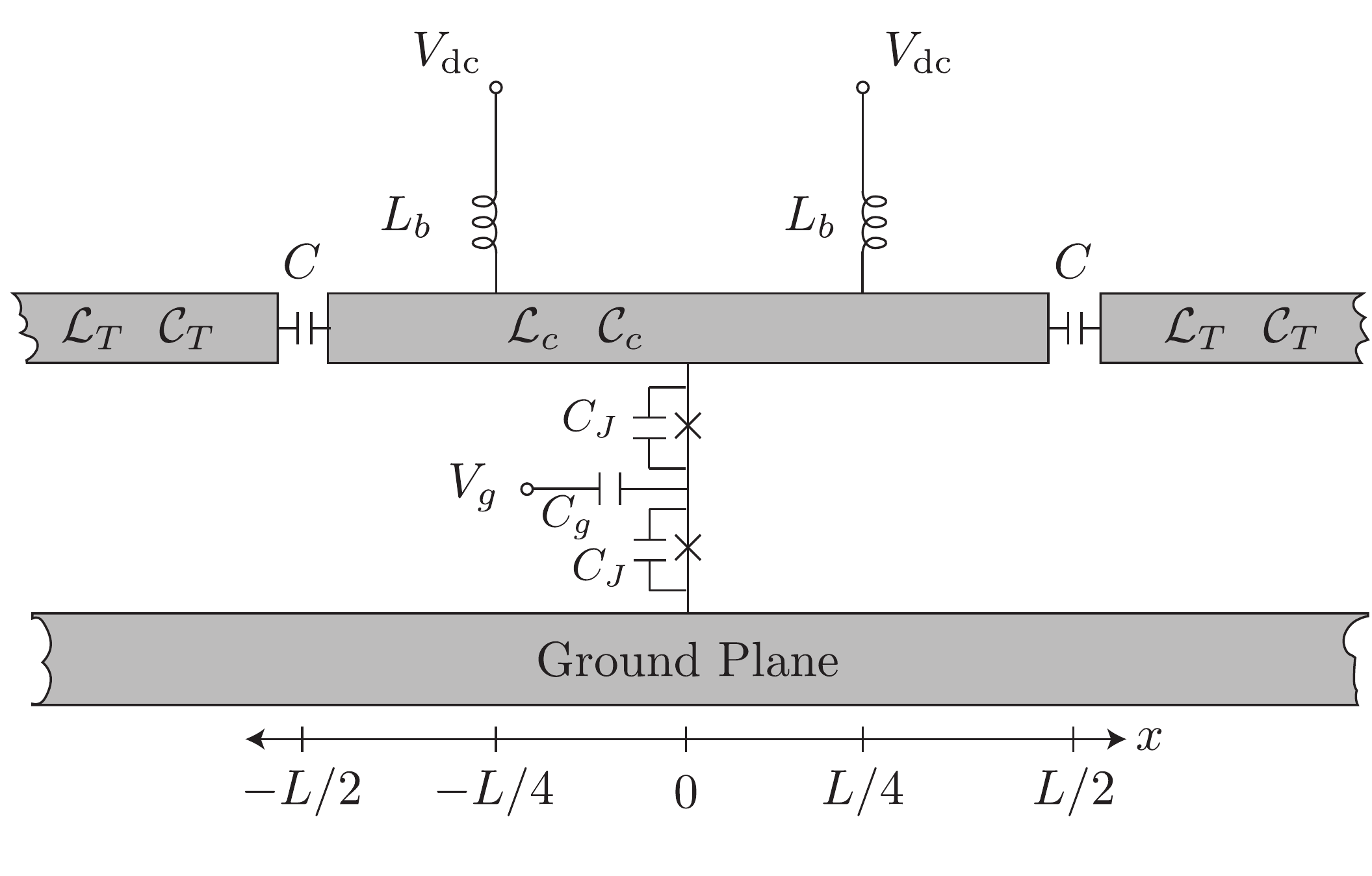}
\caption{\label{approxschemefig} Simplified model of the cavity-CPT  system, where the cavity center conductor has length $L$, and the Josephson junctions are assumed to have equal capacitances $C_J$ and critical currents $I_c$. The cavity inductance and capacitance per unit length are denoted ${\mathcal{L}_c}$, ${\mathcal{C}_c}$, respectively, while ${\mathcal{L}_T}$, ${\mathcal{C}_T}$ are the respective transmission line inductance and capacitance per unit length. The cavity and transmission line couple weakly via capacitance $C$. The dc bias lines involve inductances $L_b$. }
\end{center}
\end{figure}

In this section, we derive the cCPT Hamiltonian (1), starting with a simplified 1D model of the cCPT system shown in  Fig.~\ref{approxschemefig}. While such a model does not yield the actual mode spectrum, it does serve as a useful `scaffold' for deriving the approximate discrete mode equations and associated lumped element model, where the element parameters (i.e., capacitances, inductances etc.) can be selected so that the resonant mode frequencies and coupling strengths accurately match the experimentally determined ones. Referring to the circuit in Fig.~\ref{approxschemefig}, Kirchhoff's laws  yield the following equations in terms of the CPT phases $\gamma_{\pm}(t)=(\varphi_1(t)\pm\varphi_2(t))/2$ (with $\varphi_1$, $\varphi_2$ the gauge invariant phases across the Josephson junctions), the cavity phase field $\phi_c(x,t)$, and transmission line phase field $\phi_T(x,t)$:
\begin{widetext}
\begin{equation}
2 C_J \frac{\Phi_0}{2\pi}\frac{d^2\gamma_+}{dt^2} +2 I_c \sin\gamma_+\cos\gamma_- -C_g \frac{d V_g}{dt}=\frac{\Phi_0}{\pi {\mathcal{L}}_c} \left(\left.\frac{\partial \phi_c}{\partial x}\right|_{x=0^+}-\left.\frac{\partial \phi_c}{\partial x}\right|_{x=0^-} \right),
\label{gamma+eq}
\end{equation}
\begin{equation}
2 C_J \frac{\Phi_0}{2\pi}\frac{d^2\gamma_-}{dt^2} +2 I_c \cos\gamma_+\sin\gamma_- +C_g \frac{d V_g}{dt}=0,
\label{gamma-eq}
\end{equation}
\begin{equation}
\frac{\partial^2 \phi_c}{\partial t^2}=({\mathcal{L}}_c{\mathcal{C}}_c)^{-1}\frac{\partial^2\phi_c}{\partial x^2},~-L/2<x<0{;}~0<x<L/2,
\label{cavitywaveeq}
\end{equation}
\begin{equation}
\frac{\partial^2 \phi_T}{\partial t^2}=({\mathcal{L}}_T{\mathcal{C}}_T)^{-1}\frac{\partial^2\phi_T}{\partial x^2},~x<-L/2{;}~L/2<x,
\label{transmissionwaveeq}
\end{equation}
and junction conditions
\begin{equation}
2\frac{d\gamma_+}{d t}=\left.\frac{\partial\phi_c}{\partial t}\right|_{x=0},
\label{x=0junctioneq}
\end{equation}
\begin{equation}
\phi_c(L/4,t)-\frac{L_b}{{\mathcal{L}}_c}\left(\left.\frac{\partial\phi_c}{\partial x}\right|_{x=L^+/4}-\left.\frac{\partial\phi_c}{\partial x}\right|_{x=L^-/4}\right)=\frac{2\pi}{\Phi_0}V_{\mathrm{dc}}t,
\label{biasjunctioneq}
\end{equation}
\begin{eqnarray}
\left.-\frac{1}{{\mathcal{L}}_c}\frac{\partial\phi_c}{\partial x}\right|_{x=L^-/2}&=&\left. C\left(\frac{\partial^2\phi_c}{\partial t^2}-\frac{\partial^2\phi_T}{\partial t^2}\right)\right|_{x=L/2},\cr
\left.-\frac{1}{{\mathcal{L}}_T}\frac{\partial\phi_T}{\partial x}\right|_{x=L^+/2}&=&\left. C\left(\frac{\partial^2\phi_c}{\partial t^2}-\frac{\partial^2\phi_T}{\partial t^2}\right)\right|_{x=L/2},
\label{x=l/2junctioneq}
\end{eqnarray}
\begin{eqnarray}
\left.-\frac{1}{{\mathcal{L}}_c}\frac{\partial\phi_c}{\partial x}\right|_{x=-L^+/2}&=&\left. C\left(\frac{\partial^2\phi_T}{\partial t^2}-\frac{\partial^2\phi_c}{\partial t^2}\right)\right|_{x=-L/2},\cr
\left.-\frac{1}{{\mathcal{L}}_T}\frac{\partial\phi_T}{\partial x}\right|_{x=-L^-/2}&=&\left. C\left(\frac{\partial^2\phi_T}{\partial t^2}-\frac{\partial^2\phi_c}{\partial t^2}\right)\right|_{x=-L/2},
\label{x=-l/2junctioneq}
\end{eqnarray}
\end{widetext}
where the phase fields and their time derivatives are continuous across the junctions, and we assume $C_J\gg C_g$. We shall work in terms of the shifted cavity field: $\tilde{\phi}_c=\phi_c -\frac{2\pi}{\Phi_0}V_{\mathrm{dc}}t$, so that
Eq.~(\ref{biasjunctioneq}) becomes
\begin{equation}
\phi_c(L/4,t)-\frac{L_b}{{\mathcal{L}}_c}\left(\left.\frac{\partial\phi_c}{\partial x}\right|_{x=L^+/4}-\left.\frac{\partial\phi_c}{\partial x}\right|_{x=L^-/4}\right)=0
\label{biasjunction2eq}
\end{equation}
and  Eq.~(\ref{x=0junctioneq}) becomes
\begin{equation}
\gamma_+(t)=\phi_c(0,t)/2 +\omega_d t,
\label{x=0junction2eq}
\end{equation}
where the driving frequency is $\omega_d=\pi V_{\mathrm{dc}}/\Phi_0 =e  V_{\mathrm{dc}}/\hbar$ and we have dropped the tilde on $\phi_c$ for notational convenience. Note that we make no distinction between \vdc\ and \vsd\ in our model as we consider a simplified system in which the additional  impedance on the bias line, $Z_b$, is neglected. We can now use Eq.~(\ref{x=0junction2eq}) to eliminate $\gamma_+$ from the dynamical equations; Eqs.~(\ref{gamma-eq}) and (\ref{gamma+eq}) become respectively
\begin{widetext}
\begin{equation}
2 C_J \frac{\Phi_0}{2\pi}\frac{d^2\gamma_-}{dt^2} +2 I_c \cos\left[\phi_c(0,t)/2 +\omega_d t\right]\sin\gamma_- +C_g \frac{d V_g}{dt}=0
\label{gamma-2eq}
\end{equation}
and
\begin{equation}
\phi'_c(0,t)-\frac{C_J}{4{\mathcal{C}_c}}\phi''_c(0,t)=\frac{\pi {\mathcal{L}_c}I_c}{\Phi_0}\sin\left[\phi_c(0,t)/2+\omega_d t\right]\cos\gamma_- -\frac{\pi {\mathcal{L}_c}}{2\Phi_0} C_g\dot{V}_g,
\label{x=0boundaryeq}
\end{equation}
\end{widetext}
where we have used the cavity wave equation to replace $\ddot{\phi}_c$ with ${\phi}''_c$, and have also used $\phi'_c(0^+,t)=-\phi'_c(0^-,t)$ a consequence of the fact that only the symmetric about the origin component of the phase field $\phi_c(x,t)$ (i.e., voltage antinode/current node)  couples to the CPT, assuming idealized perfect symmetry of the device, with the `$+$' superscript on the $0$ in $\phi'_c(0,t)$ dropped for notational convenience. With this symmetry consideration,  the original length $L$ cavity with the CPT situated at the midpoint is effectively replaced with a length $L/2$ cavity and where the interaction with the CPT is expressed by  Eq.~(\ref{x=0boundaryeq}), which can be interpreted as a (rather nontrivial) boundary condition on the cavity field $\phi_c(x,t)$ at the $x=0$ end.

From now on, we set the cavity-transmission line coupling capacitance $C=0$, deriving the closed cCPT system dynamics only, described by the  $\phi_c(x,t)$ cavity wave equation~(\ref{cavitywaveeq}) for $0<x<L/2$, the coupled $\gamma_-(t)$ equation~(\ref{gamma-2eq}), the junction condition~(\ref{biasjunction2eq}) at $x=L/4$,  the boundary condition~(\ref{x=0boundaryeq}) at $x=0$, and the boundary condition
\begin{equation}
\left.\frac{\partial\phi_c}{\partial x}\right|_{x=L/2}=0
\label{x=l/2junctioneq2}
\end{equation}
at  $x=L/2$.  We solve these equations using the approximate eigenfunction expansion method, with Eq.~(\ref{x=0boundaryeq}) replaced by  the following simpler boundary condition at $x=0$:
\begin{equation}
\phi'_c(0,t)-\frac{C_J}{4{\mathcal{C}_c}}\phi''_c(0,t)\approx\left.\phi'_c(x,t)\right|_{x=-C_J/(4{\mathcal{C}}_c)}=0,
\label{x=0boundary2eq}
\end{equation}
expressed approximately as a Neumann boundary condition evaluated at the slightly shifted endpoint $x=-C_J/(4{\mathcal{C}}_c)$, with $C_J/({\mathcal{C}}_c L)\ll 1$.

We can now apply the method of separation of variables to the cavity wave equation~(\ref{cavitywaveeq}) for $0<x<L/2$: the homogeneous boundary conditions~ (\ref{x=l/2junctioneq2}), (\ref{x=0boundary2eq}) and junction condition~(\ref{biasjunction2eq}) define a Sturm-Liouville problem. Taking advantage of the smallness of the capacitance ratio $C_J/({\mathcal{C}}_c L)$, as well as the smallness of the inductance ratio, ${\mathcal{L}_c}L/L_b\ll 1$, the orthogonal eigenfunctions are approximately
\begin{widetext}
\begin{equation}
\phi_n(x)=\left\{\begin{array}{ll}
\cos \left[k_n\left(x+C_J/(4{\mathcal{C}}_c)\right)\right] & \mbox{if $ 0<x<L/4$};\\
\frac{\cos \left[k_n\left(L/4+C_J/(4{\mathcal{C}}_c)\right)\right] }{\cos \left(k_n L/4\right)}   \cos\left[k_n\left(x-L/2\right)\right]&\mbox{if $L/4<x<L/2$},\end{array}\right.
\label{eigenfunctioneq}
\end{equation}
\end{widetext}
the approximate associated wavenumber eigenvalues are
\begin{equation}
k_n=\frac{2\pi n}{L}-\frac{4\pi n}{L}\left(\frac{C_J}{4{\mathcal{C}}_c L}\right)+\left\{\begin{array}{ll}0&\mbox{if $n$ odd};\\
\frac{1}{\pi n}\frac{{\mathcal{L}}_c}{L_b}&\mbox{if $n$ even},\end{array}\right.
\label{eigenvalueeq}
\end{equation}
and the orthogonality condition on the eigenfunctions is approximately
\begin{equation}
\int_{-C_J/(4{\mathcal{C}}_c)}^{L/2} dx\ \phi_n(x)\phi_m(x)=0,\ m\neq n.
\label{orthoconditioneq}
\end{equation}

From Eqs.~(\ref{eigenfunctioneq}) and (\ref{eigenvalueeq}), we see that the odd eigenfunctions have a voltage node at approximately  $x=L/4$, while from Eq.~(\ref{biasjunction2eq}), the ratio of the ac current entering the cavity through the bias line to the average ac current flowing along the cavity center conductor at $x=L/4$ is of order $({\mathcal{L}}_c L/L_b)(C_J/{\mathcal{C}}_c L)$, i.e., second order in smallness. On the other hand, the even eigenfunctions have a voltage antinode at approximately $x=L/4$ and the ratio of the entering ac bias current to the average ac cavity current flow  is of order $({\mathcal{L}}_c L/L_b)({\mathcal{C}}_c L/C_J)\sim 1$. Thus, wave solutions involving odd eigenfunctions are expected to have low loss, while even eigenfunction wave solutions are expected to be lossy due to the presence of the bias line.

\begin{figure}
\begin{center}
\includegraphics[width=9cm]{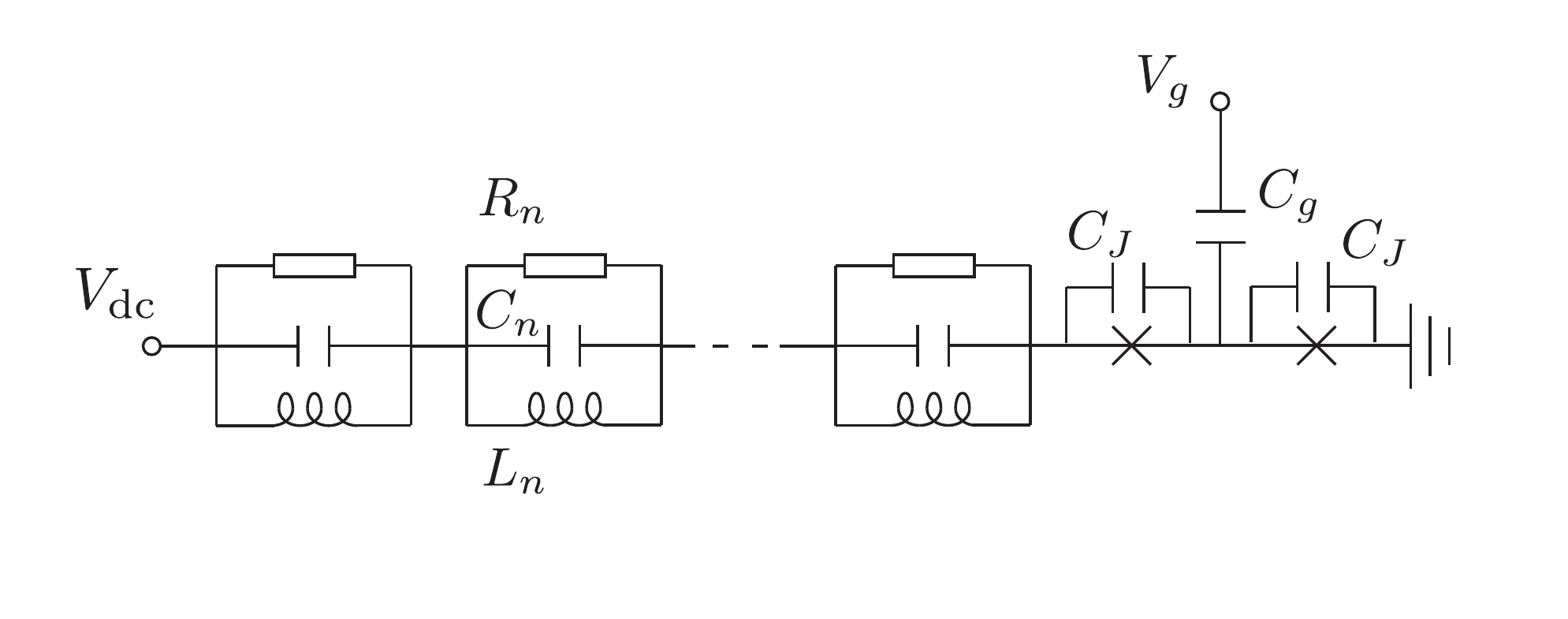}
\caption{\label{circuitfig} Lumped element circuit model of the CPT-cavity system. }
\end{center}
\end{figure}

We now assume that solutions $\phi_c(x,t)$ to the wave equation~(\ref{cavitywaveeq}) for $0<x<L/2$ with the full  boundary conditions~(\ref{x=0boundaryeq}) and (\ref{x=l/2junctioneq2}) at $x=0$ and $x=L/2$, respectively, can be expressed as a series expansion in terms of the eigenfunctions $\phi_n(x)$:
\begin{equation}
\phi_c(x,t)=\sum_n q_n(t)\phi_n(x).
\label{expansioneq}
\end{equation}
We shall assume that this series expansion can be restricted to only the odd integer eigenfunctions, with the even integer components simply accounted for through their possible effect  of  additional damping on the former. From Eq.~(\ref{expansioneq}) and the orthogonality condition~(\ref{orthoconditioneq}), the to be determined time-dependent coefficients $q_n(t)$ (for odd $n$) are given as
\begin{equation}
q_n(t)=\frac{4}{L}\int_{-C_J/(4{\mathcal{C}}_c)}^{L/2} dx\ \phi_c(x,t) \phi_n(x),
\label{coefficienteq}
\end{equation}
where the prefactor is the approximate normalization constant [neglecting $C_J/{\mathcal{C}}_c$ corrections]. Differentiating~(\ref{coefficienteq}) twice with respect to time and applying the cavity wave equation~(\ref{cavitywaveeq}), we have:
\begin{equation}
\ddot{q}_n(t)=\frac{4}{{\mathcal{L}}_c{\mathcal{C}}_c L}\int_{-C_J/(4{\mathcal{C}}_c)}^{L/2} dx\ \phi''_c(x,t) \phi_n(x).
\label{coefficient2eq}
\end{equation}
Integrating~(\ref{coefficient2eq}) by parts twice and applying the eigenvalue equation $\phi_n''(x)=-k_n^2 \phi_n(x)$ and also Eq.~(\ref{coefficienteq}), we obtain
\begin{widetext}
\begin{equation}
\ddot{q}_n(t)=-\frac{k_n^2}{{\mathcal{L}}_c{\mathcal{C}}_c} q_n(t) +\left.\frac{4}{{\mathcal{L}}_c{\mathcal{C}}_c L}\phi'_c(x,t)\phi_n(x)\right|_{-C_J/(4{\mathcal{C}}_c)}^{L/2}.
\label{coefficient3eq}
\end{equation}
Using the boundary conditions~(\ref{x=0boundaryeq}) and (\ref{x=l/2junctioneq2}), Eq.~(\ref{coefficient3eq}) becomes
\begin{equation}
\ddot{q}_n(t)=-\omega^2_n q_n(t)-\frac{4\pi I_c}{\Phi_0{\mathcal{C}}_c L} \sin\left[\frac{1}{2}\sum_{n'} q_{n'} (t)+\omega_d t\right]\cos\gamma_- +\frac{2\pi C_g \dot{V}_g}{\Phi_0{\mathcal{C}}_c L},
\label{coefficient4eq}
\end{equation}
where the free cavity mode oscillator frequencies are
\begin{equation}
\omega_n^2=\frac{k_n^2}{{\mathcal{L}}_c{\mathcal{C}}_c},
\label{modefrequencyeq}
\end{equation}
with $k_n= 2\pi n/L$, $n$ odd.
In terms of the cavity mode phase coordinates $q_n(t)$, the $\gamma_-$ equation~(\ref{gamma-2eq}) becomes
\begin{equation}
2 C_J \frac{\Phi_0}{2\pi}\frac{d^2\gamma_-}{dt^2} +2 I_c \cos\left[\frac{1}{2}\sum_{n} q_{n} (t)+\omega_d t\right]\sin\gamma_- +C_g \frac{d V_g}{dt}=0.
\label{gamma-3eq}
\end{equation}
\end{widetext}
Equations~(\ref{coefficient4eq}) and (\ref{gamma-3eq}) give the approximate discrete mode description of the closed cCPT system dynamics. In modeling the experiment, the various circuit parameters appearing in (\ref{coefficient4eq}) and (\ref{gamma-3eq}) can be selected so as to provide the best fit to the data. In this way, the effective discrete mode equations are assumed to be more versatile than the original starting equations at the beginning of this section, which are tied to a particular model of the cavity geometry.
Figure~\ref{circuitfig} shows a lumped element circuit model that yields the above discrete mode equations (neglecting small capacitance ratio terms), where the lumped capacitance and inductance elements are $C_n={\mathcal{C}}_c L/2$, $L_n=1/(\omega_n^2 C_n)$, respectively, and dissipative effective resistance elements $R_n$ have also been included for completeness.

The closed system equations of motion~(\ref{coefficient4eq}) and (\ref{gamma-3eq}) follow from the Hamiltonian
\begin{widetext}
\begin{eqnarray}
{\mathcal{H}}&=&\left(\frac{2\pi}{\Phi_0}\right)^2\sum_n\frac{1}{2 C_n}\left(p_n+\frac{\Phi_0}{4\pi}C_g V_g\right)^2+\left(\frac{\Phi_0}{2\pi}\right)^2\sum_n\frac{q_n^2}{2L_n} \cr
&& +4 E_c (N-n_g/2)^2 -2 E_J \cos\left[\frac{1}{2}\sum_{n} q_{n} +\omega_d t\right]\cos\gamma_-,
\label{hamiltonianeq}
\end{eqnarray}
where the mode sums are restricted to odd $n$,  $N=p_-/\hbar$ is minus the number of excess Cooper pairs on the island, $n_g=C_g V_g/e$ is the polarization charge induced by the applied gate voltage bias $V_g$, $E_c=e^2/(2C_J)$ is the approximate CPT charging energy (neglecting $C_g$), and $E_J =I_c\Phi_0/(2\pi)$ is the Josephson energy of a single JJ.

The quantum Hamiltonian corresponding to Eq.~(\ref{hamiltonianeq}) can be written as
\begin{eqnarray}
H&=&\sum_n\hbar\omega_n a_n^{\dagger}a_n +4 E_{c}\sum_{N=-\infty}^{+\infty}\left(N-n_g/2\right)^2|N\rangle\langle N|\cr
&&-E_J\sum_{N=-\infty}^{+\infty}\left(|N+1\rangle\langle N|+|N-1\rangle\langle N|\right)\cos\left[\sum_n\Delta_n (a_n+a_n^{\dagger}) +\omega_d t\right],\label{quantumhamiltonianeq}
\end{eqnarray}
\end{widetext}
where we have neglected the gate voltage dependent term in the cavity mode coordinate part of the Hamiltonian and where $\Delta_n$ is the zero-point uncertainty of the cavity mode phase  coordinate $q_n$:
\begin{equation}
\Delta_n=\sqrt{\frac{\pi\sqrt{L_n/C_n}}{R_K}}=\sqrt{\frac{Z_n}{R_K}},
\label{zeropointeq}
\end{equation}
with $Z_n$ the cavity mode impedance and $R_K=h/e^2\approx 25.8~{\mathrm{k}}\Omega$ the von Klitzing constant. Restricting to the lowest, $n=1$ cavity mode and truncating  to a two-dimensional subspace involving linear combinations of only zero ($|0\rangle$) and one ($|1\rangle$) excess Cooper pairs on the island then yields the cCPT Hamiltonian (1) given in the main text.

\subsection{Quantum dynamics of the model system}
In this Section we investigate the dynamics of the single-mode system within the charge state basis $|0\rangle$ and $|1\rangle$.
Generically the dynamics of the system is described by a master equation of the form,
\begin{equation}
\dot{\rho}=-\frac{i}{\hbar}[H,\rho]+\mathcal{L}_N\rho+\mathcal{L}_n\rho, \label{eq:me}
\end{equation}
where $H$ is given by (1), the terms $\mathcal{L}_N\rho$ and $\mathcal{L}_n\rho$ describe the effect of the environment on the island charge and cavity mode respectively.

We will not derive the dissipative parts of the master equation here, but instead take a phenomenological approach. We assume the simplest possible form for $\mathcal{L}_N\rho$, allowing transitions between charge states (for $\varepsilon>0$),
\begin{equation}
\mathcal{L}_N\rho={\Gamma}\left(\sigma_-\rho\sigma_+-\frac{1}{2}\left\{\sigma_+\sigma_-,\rho\right\}\right),
\end{equation}
where $\Gamma$ is the charge relaxation rate and $\sigma_+=(\sigma_-)^{\dagger}=|1\rangle\langle 0|$, an approach which we expect to provide a reasonable description of the dissipative dynamics in the regime where $n_g\ll 1$.
For the cavity mode we assume dissipation due to an oscillator bath at zero temperature and hence we have,
\begin{equation}
\mathcal{L}_n\rho=\frac{\omega_0}{Q}\left(a\rho a^{\dagger}-\frac{1}{2}\left\{a^{\dagger}a,\rho\right\}\right),
\end{equation}
where $Q$ is the quality factor.

\begin{widetext}
\subsubsection{Semi-classical description}

Given the master equation, Eq.\ (\ref{eq:me}), we can write down equations of motion for the expectation values of the operators, $a$, $\sigma_+$ and $\sigma_z$,

\begin{eqnarray}
\langle\dot a\rangle&=&-\left(i\omega_0+\frac{1}{2Q}\right)\langle a\rangle-i\frac{E_J\Delta}{\hbar}\langle\sin[\Delta(a+a^{\dagger})+\omega_dt](\sigma_++\sigma_-)\rangle\\
\langle\dot{\sigma}_+\rangle&=&\left(i\frac{2\varepsilon}{\hbar}-\Gamma\right)\langle\sigma_+\rangle+i\frac{E_J}{\hbar}\langle\cos[\Delta(a+a^{\dagger})+\omega_dt]\sigma_z\rangle\\
\langle\dot{\sigma}_z\rangle&=&i\frac{2E_J}{\hbar}\langle\cos[\Delta(a+a^{\dagger})+\omega_dt](\sigma_+-\sigma_-)\rangle-\Gamma(\langle \sigma_z\rangle+1).
\end{eqnarray}
We proceed by making a semi-classical approximation\,\cite{Walls:2008}, treating expectation values of products of operators as products of expectation values.
Adopting the notation $\alpha=\langle a\rangle$, $\sigma=\langle\sigma_+\rangle$, $z=\langle\sigma_z\rangle$, we obtain the following self-contained equations of motion for the system,
\begin{eqnarray}
\dot{\alpha}&=&-\left(i\omega_0+\frac{1}{2Q}\right)\alpha-i\frac{E_J\Delta}{\hbar}\sin[\Delta(\alpha+\alpha^*)+\omega_dt](\sigma+\sigma^*)\label{eq:mf1}\\
\dot{\sigma}&=&\left(i\frac{2\varepsilon}{\hbar}-\Gamma\right)\sigma+i\frac{E_J}{\hbar}\cos[\Delta(\alpha+\alpha^*)+\omega_dt]z \label{eq:mf2}\\
\dot{z}&=&i\frac{2E_J}{\hbar}\cos[\Delta(\alpha+\alpha^*)+\omega_dt](\sigma-\sigma^*)-\Gamma(z+1). \label{eq:mf3}
\end{eqnarray}
\end{widetext}

\subsubsection{Origin of the resonances}
The sequential tunneling resonances shown schematically in Fig.~1e (left panel) involve a two step cycle of processes which transfer Cooper pairs across one junction at a time. In the first step, a Cooper pair tunnels across one of the junctions taking the island from the lower to the higher energy charge state whilst at the same time a photon is emitted into the cavity. This process occurs when the voltage lost by the Cooper pair tunneling across the junction balances the change in charging energy and the cost of photon production, $\hbar\omega_d=2|\varepsilon|+\hbar\omega_0$, which defines the diagonal resonance lines. After this first step, the island is in the higher energy charge state so the system can return to the lower charge state in the second step via the tunneling of a Cooper pair across the other junction (with the extra energy being dissipated not in the cavity mode, but in other environmental degrees of freedom, i.e. through the $\mathcal{L}_N\rho$ terms in the master equation).

The simplest way of understanding the locations of the cotunneling resonances shown in Fig.~1e (right panel) is in terms of matching between the voltage energy lost by a Cooper pair traversing both CPT junctions and the energy required to create $k$ photons, $2e\vdc=2\hbar\omega_d=k\hbar\omega_0$. As one would expect, the cotunneling resonances appear in the semi-classical equations. Their locations can be determined (though not what the corresponding steady states are) by assuming that the occupation of the cavity is small (i.e.\ $\alpha\ll 1$) so that the effect of the cavity on the dynamics of the CPT island charge can be neglected. Dropping the dependence on $\alpha$ in Eqs.\ (\ref{eq:mf2}) and (\ref{eq:mf3}), we are left with a pair of equations which describes a two-level system subject to both a harmonic drive and damping,
\begin{eqnarray}
\dot{\sigma}&=&\left(i\frac{2\varepsilon}{\hbar}-\Gamma\right)\sigma+i\frac{E_J}{\hbar}\cos(\omega_dt)z \label{eq:mf2b}\\
\dot{z}&=&i\frac{2E_J}{\hbar}\cos(\omega_dt)(\sigma-\sigma^*)-\Gamma(z+1). \label{eq:mf3b}
\end{eqnarray}
The long time behavior of the CPT island charge will be a periodic function of the drive frequency. For reasons of symmetry\,\cite{Grifoni:1998}, $\sigma(t)$ contains only odd harmonics  and hence can be expressed as the Fourier series
\begin{equation}\label{fseq}
\sigma(t)=\sum_n' c_{n}{\rm e}^{in\omega_d t},
\end{equation}
with the prime indicating that the sum runs over odd integers.

Defining $\tilde{\alpha}=\alpha{\rm e}^{i\omega_0 t}$, and substituting the Fourier series (\ref{fseq}) for $\sigma(t)$,  into the Eq.\ (\ref{eq:mf1}) we find
\begin{eqnarray}
\dot{\tilde{\alpha}}&=&-\frac{\tilde{\alpha}}{2Q}-i\frac{E_J\Delta}{\hbar}{\rm e}^{i\omega_0 t}\sum'_n\left(c_{n}{\rm e}^{in\omega_d t}+c^*_{n}{\rm e}^{-in\omega_d t}\right) \nonumber \\
&&\times\sin\left[\Delta\left(\tilde{\alpha}{\rm e}^{-i\omega_0 t}+\tilde{\alpha}^*{\rm e}^{i\omega_0 t}\right)+\omega_dt\right]. \label{eq:alpha}
\end{eqnarray}
Resonances occur whenever the damping term is matched by another term which has no explicit time dependence. Expanding out the sinusoidal term, we see  that there are exponentials with all integer powers of $\omega_0$, but only even powers of $\omega_d$, hence the most general resonance condition is  $k\omega_0=2p\omega_d$, corresponding to $k$ photons being produced by the cotunneling of $p$ Cooper pairs. In addition there is a time independent contribution which is present for any combination of $\omega_d$ and $\omega_0$ (it is generated by terms in the expansion of the sine that are proportional to ${\rm e}^{-i\omega_0 t}$). However, this term is dispersive leading to a frequency shift in the cavity mode and it does not affect the energy.

At the one photon resonance where $\omega_d=\omega_0/2$, oscillations of $\sigma(t)$ at frequency $\omega_d$ act to upconvert the drive oscillations to produce oscillations at frequency $2\omega_d$ which are resonant with the cavity mode. Expanding Eq.\ (\ref{eq:alpha}) to linear order in $\tilde{\alpha}$ and retaining only terms without explicit time dependence we find,
\begin{eqnarray}
\dot{\tilde{\alpha}}&=&-\frac{\tilde{\alpha}}{2Q}+\frac{E_J\Delta}{2\hbar}\beta_0-i\frac{E_J\Delta^2}{2\hbar}\left(\beta_1\tilde{\alpha}+\beta_2\tilde{\alpha}^*\right)
\label{eq:1phr}
\end{eqnarray}
where the $\beta$ coefficients depend on the Fourier components, 
\begin{eqnarray}
\beta_0&=&c_{-1}+c_{1}^*-c_{-3}-c_{3}^*\\
\beta_1&=&c_{-1}+c_{1}+c_{-1}^*+c_{1}^*\\
\beta_2&=&c_{-3}+c_{3}^*+c_{-5}+c_{5}^*.
\end{eqnarray}
The $\beta_0$ and $\beta_2$ terms in Eq.\ (\ref{eq:1phr}) arise because of the resonance condition $\omega_d=\omega_0/2$ and  lead to changes in the energy of the mode, with the $\beta_0$ term acting like a linear drive. The $\beta_1$ term is independent of the specific choice of $\omega_d$ and generates only a frequency shift (since $\beta_1$ is real).

A similar picture emerges at the two photon resonance where $\omega_d=\omega_0$ where an expansion leads to
\begin{eqnarray}
\dot{\tilde{\alpha}}&=&-\frac{\tilde{\alpha}}{2Q}-i\frac{E_J\Delta^2}{2\hbar}\left(\beta_1\tilde{\alpha}+\beta_3\tilde{\alpha}^*\right),
\label{eq:2phr}
\end{eqnarray}
where here $\beta_3=c_{-3}+c_{3}^*+c_{-1}+c_{1}^*$. The $\beta_1$ term again leads to a frequency shift while the $\beta_2$ term, which arises because of resonance condition ($\omega_d=\omega_0$), generates changes in the energy of the mode.

 The simple resonance condition suggests that, in a crude sense, the CPT can be thought of as acting like a single effective Josephson junction at the cotunneling resonances as there is no explicit energy matching condition involving the internal state of the island (and hence the gate voltage), though the behavior of the system {\it does} depend on $n_g$ implicitly through the dynamics of $\sigma$. However, at points where the two island charge states are degenerate (i.e.\ $n_g=1$ and hence $\varepsilon=0$) the driven oscillations in $\sigma$ described by Eqs.\ (\ref{eq:mf2}) and (\ref{eq:mf3}) become entirely imaginary leading to a decoupling from the cavity (since it is the combination $\sigma+\sigma^*$ which appears in Eq.\ \ref{eq:mf1} for $\alpha$), a feature which should not depend on the details of the dissipative terms. In Eqs.\ (\ref{eq:1phr}) and  (\ref{eq:2phr}) the $\beta$ coefficients vanish for a pure imaginary $\sigma$ since in that case $c_{n}=-c_{-n}^*$.  Thus both the driving of the cavity and the frequency shift terms generated by the interaction with the CPT should shut off at the charge degeneracy point. This provides an explanation for the loss of cavity output seen around $n_g=\pm 1$ in the experiment.

\subsubsection{Dynamics at the cotunneling resonances}

\begin{figure}[t]
\centering
{\includegraphics[width=8.5cm]{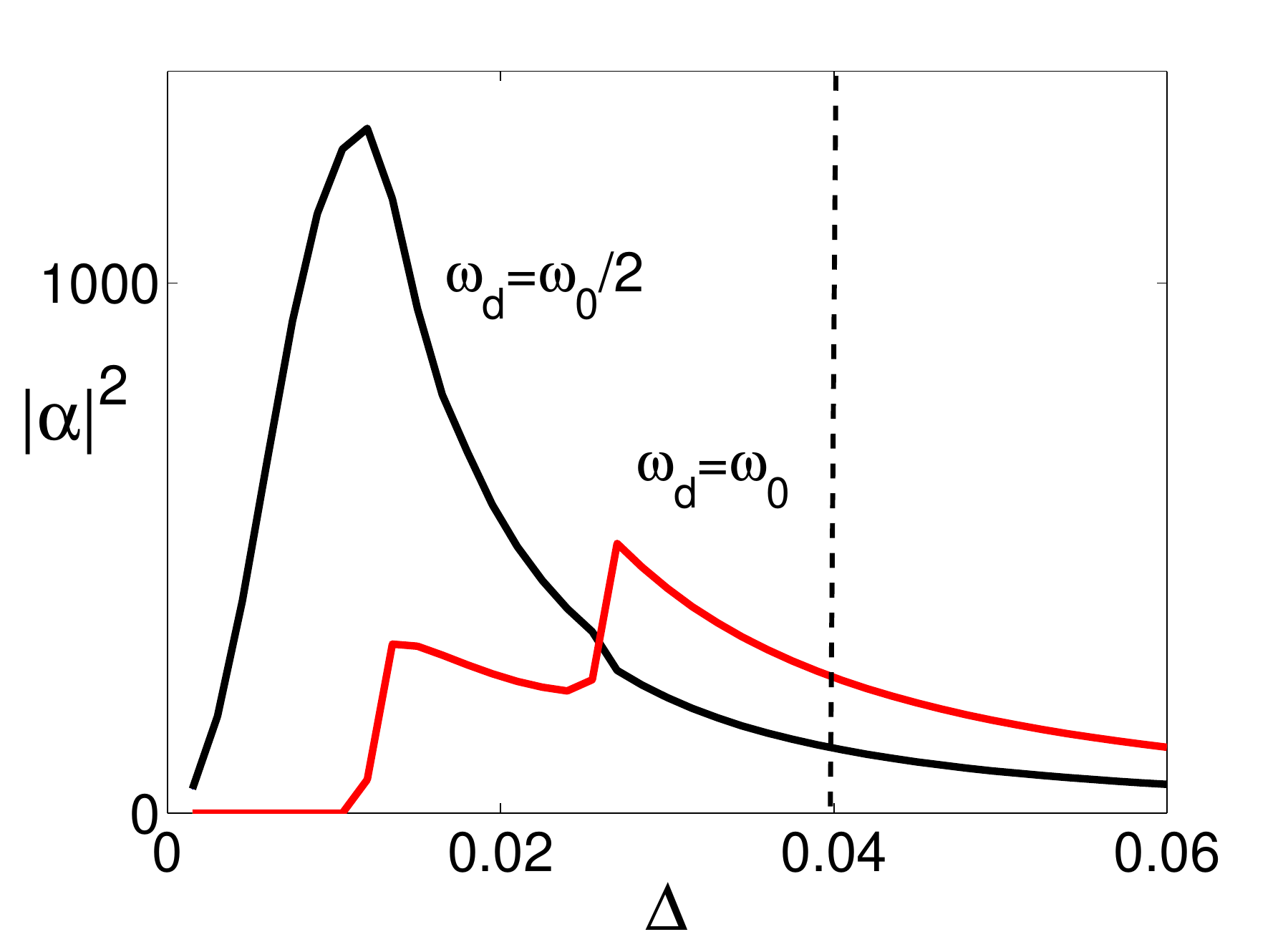}
}
\caption{Average cavity energy calculated semi-classically, $|\alpha|^2$, in the long time limit at the one and two photon resonances as a function of $\Delta$. We use the parameter values from the experiment  $Q=3500$, $E_J/\hbar\omega_0=3.2$, $4E_c/\hbar\omega_0=6.6$, $n_g=0$ and choose $\Gamma/\omega_0=0.02$.}
\label{fig3}
\end{figure}

\begin{figure}[t]
\centering
{\includegraphics[width=8.5cm]{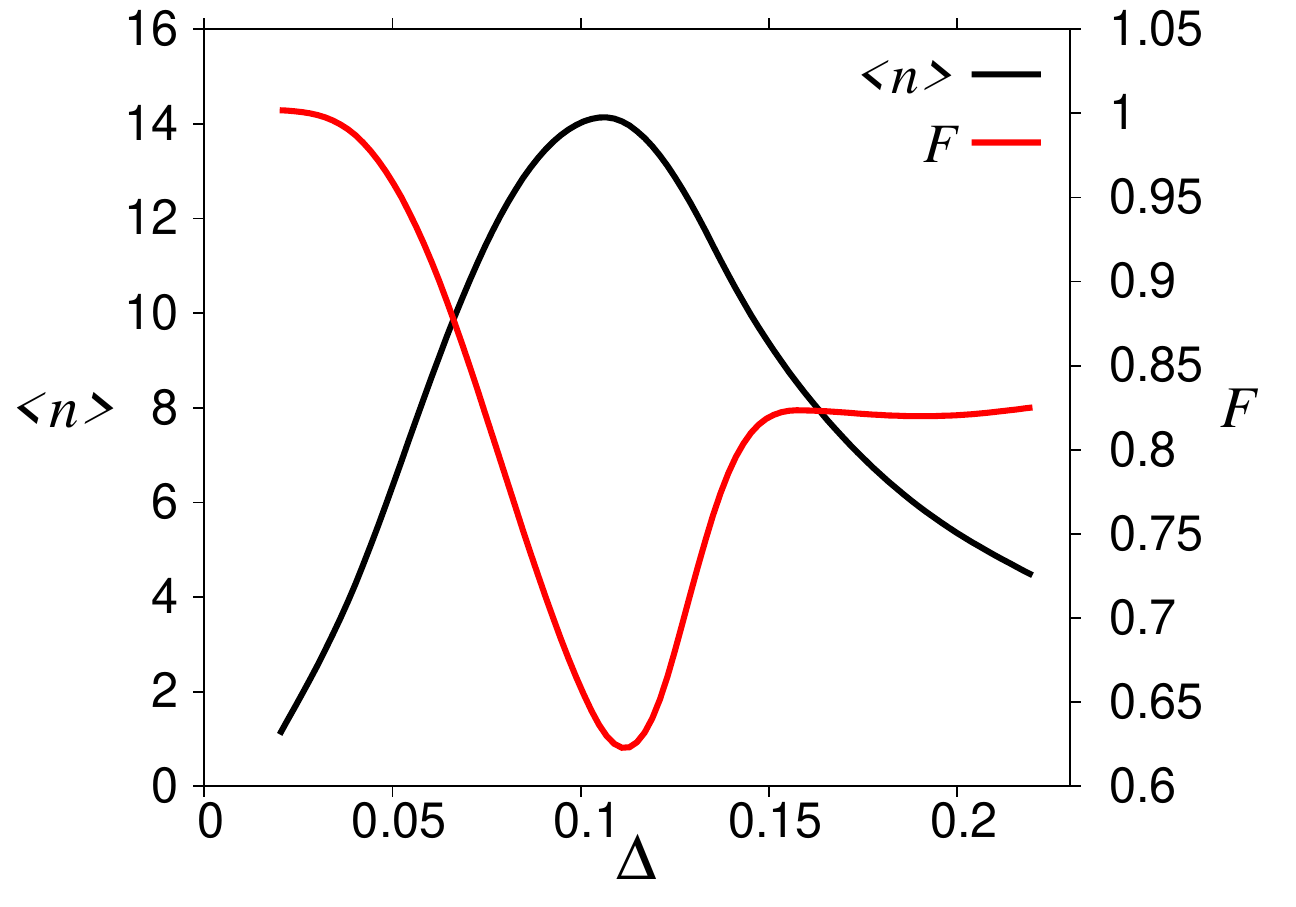}
}
\caption{Average cavity occupation, $\langle n\rangle$, and Fano factor, $F$, as a function of $\Delta$ for $\omega_d=\omega_0/2$ in the limit of long times. Here $E_J/\hbar\omega_0=2$, $\varepsilon/\hbar\omega_0=2.5$, $Q=75$ and $\Gamma/\omega_0=0.2$.}
\label{fig4}
\end{figure}

The level of cavity excitation predicted by the model can be obtained by integrating Eqs.\ (\ref{eq:mf1})-(\ref{eq:mf3}) and calculating the long time average of $|\alpha|^2$ which is equivalent to the average number of cavity photons. Figure \ref{fig3} shows an example of the behavior as a function of $\Delta$ at the one and two photon resonances. For the one photon resonance there is a smooth increase in energy as a function of $\Delta$ followed by a peak and then a decaying region. The two photon resonance grows abruptly in two stages before also decaying in a similar way to the one photon case. For the experimental value $\Delta=0.04$ we obtain photon numbers of $122$ for $\omega_d=\omega_0/2$ and $256$ for $\omega_d=\omega_0$, which are very similar in magnitude to the results described in the main text.

Numerical integration of the master equation (\ref{eq:me}) allows us to examine the quantum fluctuations in the system\,\cite{Johansson:2012}, although we have not explored the regime probed in the experiment as the photon numbers involved are too high (instead larger values of $\Delta$ are combined with much lower values of $Q$). Such calculations certainly suggest that amplitude-squeezing  ($F<1$) is a generic property of the system. Figure \ref{fig4} shows an example of the kind of results which can be obtained in this way for $\omega_d=\omega_0/2$. The average energy behaves in the same general way as that obtained from the semi-classical equations for the experimentally relevant parameters (shown in Fig.\ \ref{fig3}). Amplitude-squeezing occurs over the whole range of $\Delta$ studied. 

Further evidence of the system's very general tendency to display amplitude-squeezing comes from analysis of a model  describing an analogous one junction system\,\cite{Armour:2013}. In this system Fano factors less than unity are found generically for a very broad range of parameters at the one and two photon resonances, including the regime where $Q\ge 1000$ and $\Delta\simeq 0.04$. Finally, we point out that the lowest Fano factors are usually associated with systems where there is an interaction between a harmonic mode and a two-level system, a prominent example being the trapping states of the micromaser\,\cite{Haroche:2006} (where almost pure Fock states can in principle be generated). Indeed, recent work by Marthaler et al.\,\cite{Marthaler:2011} predicted the existence of trapping states in a model CPT-cavity system for processes where changes in island charge state are associated with photon emission (closely related to the sequential resonances observed in the experiment).


\end{document}